\documentclass[aps,prx,twocolumn,amsfonts,superscriptaddress,showpacs,floatfix,notitlepage]{revtex4-1}
\pdfoutput=1
\usepackage{comment}
\usepackage{enumerate}
\usepackage{amssymb}
\usepackage{amsmath,bm,bbm}
\usepackage{braket}
\usepackage{amsthm}
\usepackage{graphicx}
\usepackage{mathtools}
\usepackage[usenames,dvipsnames]{color}
\usepackage[colorlinks,bookmarks=false,citecolor=NavyBlue,linkcolor=OliveGreen,url
color=blue]{hyperref}
\usepackage{tensor}
\newcommand{\be}{\begin{equation}}
\newcommand{\ee}{\end{equation}}
\newcommand{\ba}{\begin{aligned}}
\newcommand{\ea}{\end{aligned}}
\newcommand{\1}{\mathbbm{1}}
\newcommand{\one}{\mathbbm{1}}

\newtheorem{property}{Property}
\theoremstyle{break}        

\newtheorem{lemma}{Lemma}

\begin{document}
\title{Entanglement spreading in a minimal model of maximal many-body quantum chaos}
\author{Bruno Bertini, Pavel Kos, and Toma\v z Prosen}

\affiliation{Department of Physics, Faculty of Mathematics and Physics, University of Ljubljana, Jadranska 19, SI-1000 Ljubljana, Slovenia}

\begin{abstract}
The spreading of entanglement in out-of-equilibrium quantum systems is currently at the centre of intense interdisciplinary research efforts involving communities with interests ranging from holography to quantum information. Here we provide a constructive and mathematically rigorous method to compute the entanglement dynamics in a class of ``maximally chaotic", periodically driven, quantum spin chains. Specifically, we consider the so called ``self-dual" kicked Ising chains initialised in a class of separable states and devise a method to compute exactly the time evolution of the entanglement entropies of finite blocks of spins in the thermodynamic limit. Remarkably, these exact results are obtained despite the models considered are maximally chaotic: their spectral correlations are described by the circular orthogonal ensemble of random matrices on all scales. Our results saturate the so called ``minimal cut" bound and are in agreement with those found in the contexts of random unitary circuits with infinite-dimensional local Hilbert space and conformal field theory. In particular, they agree with the expectations from both the quasiparticle picture, which accounts for the entanglement spreading in integrable models, and the minimal membrane picture, recently proposed to describe the entanglement growth in generic systems. Based on a novel ``duality-based" numerical method, we argue that our results describe the entanglement spreading from any product state at the leading order in time when the model is non-integrable.   
\end{abstract}

\maketitle
\tableofcontents

%%%%%%%%%%%%
\section{Introduction}%%
%%%%%%%%%%%%

Entanglement is arguably the most distinctive feature of quantum mechanics. It generates a special kind of non-local correlations which can be present in quantum states but have no analogues in the classical realm. While its elusive nature puzzled physicists for many years~\cite{EPR, Bell}, it is currently regarded as a powerful resource for advances both in technological applications and in the theoretical understanding of the physical world. In particular, it plays a crucial role in the study of quantum many-body systems out of equilibrium~\cite{Entrev1,Entrev2}. This is due to two main reasons. First, the growth of entanglement during the non-equilibrium dynamics measures the increasing complexity of a time-evolving quantum state, with immediate implications on the feasibility of tensor network simulations~\cite{SWVC:PRL, SWVC:NJP, PV, HCTDL, Dubail}. Second, the evolution of the entanglement gives crucial information on how equilibrium statistical mechanics emerges from many-body quantum dynamics. Specifically, it is now understood that the thermodynamic entropy of the statistical ensemble describing local observables at infinte times is a measure of the entanglement accumulated during the time evolution~\cite{p-18,dls-13, bam-15, Gur14, SPR11}.

Moreover, the very way in which the entanglement spreads for finite times appears to be among the most universal aspects of many-body dynamics. Consider for instance an initial separable state, where none of the local constituents is entangled with any other. Switching on spatially local Hamiltonian interactions throughout the system (a procedure called ``global quench''), one finds quite generally that the bipartite entanglement between a large connected region of the system and the rest grows linearly in time. This scenario has been observed in a large number of analytical and numerical  investigations~\cite{CC, DeChiara, leda-2014, FC:exactXY, ep-08, nr-14, bhy-17, hbmr-17, lauchli-2008, KH:NonIntEnt, fc-15, buyskikh-2016, cotler-2016, Alba, mbpc-17, mkz-17, FNR:longrangehigherd, ckt-18, BTC:beyondpairs, BC:beyondpairs2, CLM:minimalcut, LS:CFT, alba-inh, BFPC18, daley-2012, evdcz-18, PL:kickedIsing, ABGH:CFT, LM:CFT} and recently even in cold atomic experiments~\cite{expS}. Known exceptions to this empirical fact are systems exhibiting real space localization \cite{DeChiara,ZPP}, confinement \cite{KCTC:confinement}, and quenched disorder creating weak links~\cite{Nahum3:disordered}. In particular, the logarithmic spreading of entanglement in the presence of many-body localization (MBL)~\cite{MBL} is one of the main defining features of the MBL phase.

The linear growth of entanglement after a global quench has been first observed in the context of {$(1+1)$-dimensional} conformal field theory (CFT), where it has been explained in terms of an intuitive quasiparticle picture~\cite{CC}. The initial state, which is not an eigenstate of the Hamiltonian, can be thought of as a collection of pairs of oppositely moving quasiparticles. These, in the course of time, spread the entanglement across the system, in a similar way to the one conceived in the historical gedanken experiment by Einstein, Podolsky and Rosen \cite{EPR}. In this picture, the entanglement between two different portions of the system is given by the number of pairs sharing a particle with each portion. The same idea can be used to explain the entanglement spreading in systems with stable quasiparticle excitations, for instance free~\cite{FC:exactXY} and interacting~\cite{Alba} integrable models. It does not account, however, for the linear growth of entanglement observed in systems with no identifiable quasiparticle content such as generic interacting systems~\cite{KH:NonIntEnt, lauchli-2008, PL:kickedIsing} or holographic CFTs~\cite{CLM:minimalcut, LS:CFT, ABGH:CFT, LM:CFT}.

More recently, a fruitful avenue of research came from the study of the so-called random quantum circuits~\cite{RandomCircuitsEnt, r-2017, Nahum2, Nahum3:disordered, Nahum4, Nahum5, CDC1}, where the dynamics is completely random in space and the only constraint is given by the locality of interactions. In this case the linear growth of entanglement can be explained using a ``minimal membrane" picture~\cite{RandomCircuitsEnt, Nahum4}, which is conjectured to apply, at least at the qualitative level, to generic (non-integrable) clean and noisy systems in any spatial dimension. In essence one quantifies the amount of entanglement between two portions of the system by measuring the surface of the minimal space-time membrane separating the two portions. This picture has been analytically tested in certain limiting regimes of random quantum circuits, specifically assuming that the Hilbert space dimension $q$ per local constituent is large ($q\gg 1$).  Results are available both when the dynamics are random also in time~\cite{RandomCircuitsEnt, Nahum5}, and when they are periodically driven~\cite{CDC1}. No analytical result, however, exists on entanglement dynamics in specific non-integrable models with local interactions and small finite $q$ or, in general, for clean systems. 

In this paper we fill this gap providing exact results for the entanglement dynamics of quantum-chaotic spin chains with two-dimensional local Hilbert space ($q=2$). Specifically, we consider a family of Floquet-Ising chains which undergo a transition between integrability and ergodicity (or quantum chaos) by turning on a longitudinal magnetic field.  The latter may be either spatially homogeneous or arbitrarily spatially modulated. Note that the non-integrability of the model for non vanishing longitudinal magnetic fields has recently been proved by computing exactly the spectral statistics~\cite{letter}. We identify a class of separable initial states, homogeneous or arbitrarily modulated in space, from which the entanglement dynamics can be computed exactly for any non-disjoint bipartition. These results are of high significance for three main reasons. (i) They provide an exact verification of both the linear growth of entanglement and its relaxation to the thermodynamic entropy in concrete quantum-chaotic models. (ii) They provide a general method allowing one to obtain exact results for the non-equilibrium dynamics of many-body quantum systems even in the absence of integrability. (iii) They are valid in both the integrable and the non-integrable case, allowing for a unified interpretation of the entanglement spreading.

The paper is laid out as follows. In Sec.~\ref{sec:model} we present the model considered, define the protocol used to drive it out of equilibrium, and introduce the entanglement measures of interest. In Sec.~\ref{sec:results} we present a comprehensive summary and discussion of our results. In Sec.~\ref{sec:duality} we explain the duality mapping which is the key for our analytical calculations. In Sec.~\ref{sec:separating} we identify the classes of initial states leading to an exactly solvable entanglement dynamics. In Sec.~\ref{sec:specialcase} we sketch the main steps of the calculation. In Sec.~\ref{sec:genericcase} we present a thorough numerical analysis of the entanglement spreading from generic separable initial states and advocate that our exact results give the leading-order in time description of the non-integrable case. Finally, Sec.~\ref{sec:conclusions} contains our conclusions. A number of technical points and proofs are reported in the appendices.

%%%%%%%%%%%%%%%%%%%%%%%%%%%%%%%%%%%%%%%%%%%%%%%%%%%%%%%%%%
\section{Model, quench protocol, and observables of interest}%%
%%%%%%%%%%%%%%%%%%%%%%%%%%%%%%%%%%%%%%%%%%%%%%%%%%%%%%%%%%
\label{sec:model}

The main objective of this paper is to determine a minimal quantum chaotic model~\cite{note2}, with local interactions and finite local Hilbert space, allowing for an exact determination of the entanglement spreading. A candidate emerging naturally in our quest is the so called kicked Ising model~\cite{KI_JPA,KI_Ruelle,KI_PRE}, which describes a classical Ising model in the presence of a longitudinal magnetic field and periodically kicked with a transverse magnetic field. This model is quantum chaotic in the sense that its spectral statistics are described by the circular orthogonal ensemble of random matrices~\cite{PP-KI}, but, as we recently proved~\cite{letter}, at some specific points of its parameter space it allows for exact calculations. This is because at these points, called ``self-dual" points (see below), the system acquires a remarkable algebraic structure making of it a {\em maximal scrambler with local interactions}.

To be more specific, let us introduce the Hamiltonian of the kicked Ising model. Setting to one the time interval between the kicks we have 
\be
H_{\rm KI}[\boldsymbol h;t]=H_{\rm I}[\boldsymbol h]+\sum_{m=-\infty}^\infty\!\!\!\delta(t-m) H_{\rm K}\,,
\label{eq:ham}
\ee
where $\delta(t)$ is the Dirac delta function and we defined 
\begin{align}
&H_{\rm I}[\boldsymbol h]\equiv\! J \sum_{j=1}^{L} \sigma^{z}_j \sigma^z_{j+1}+ \sum_{j=1}^{L} h_j \sigma^z_{j}\,,\label{eq:hamiltonians1}\\
&H_{\rm K}\equiv b \sum_{j=1}^{L}  \sigma^x_j.
\label{eq:hamiltonians2}
\end{align}
In these equations $L$ is the volume of the system, the matrices $\sigma_j^a$, with ${a\in\{x,y,z\}}$, are the Pauli matrices at position $j$, and we use periodic boundary conditions adopting the notation convention $\sigma_{L+1}^{a}\equiv\sigma_1^{a}$. 

The parameters $J$ and $b$ are, respectively, the Ising coupling and the transverse ``kicking'' field, while the $L$-component vector $\boldsymbol h=(h_1,\ldots, h_L)$ describes a position dependent longitudinal field. As anticipated before, in this paper we consider some specific points in the parameter space of the model. In particular we focus on the ``self-dual'' points, specified by the condition  
\be
|J|=|b|= \frac{\pi}{4}.
\label{eq:selfduals}
\ee
In Secs.~\ref{sec:duality}, \ref{sec:separating}, and \ref{sec:specialcase} we explain how, at these points, a duality symmetry of the model allows for an analytical treatment of the entanglement dynamics. To be specific, from now on we set 
\be
J=  \frac{\pi}{4},\qquad\qquad b=-\frac{\pi}{4},
\label{eq:selfdual}
\ee
but our results apply to all the four combinations fulfilling \eqref{eq:selfduals}. The longitudinal magnetic fields ${\boldsymbol h}$ are instead left arbitrary and are used to switch between the integrable and the non-integrable case. Indeed, for ${\boldsymbol h = \boldsymbol 0}$ the Hamiltonian \eqref{eq:ham} is integrable (it can be mapped to a problem of non-interacting fermions), while it is ergodic (non-integrable) for a generic choice of longitudinal fields. In the latter case the only symmetry of \eqref{eq:ham} is time reversal. This is represented by the anti-unitary operator $T$, defined by its action on the spin variables as follows 
\be
\sigma_j^a\mapsto T \sigma_j^a T =\sigma_j^{a\,*}\,.
\ee
Here $(\cdot)^*$ denotes the complex conjugation in the ``computational'' basis, composed by simultaneous eigenstates of $\{\sigma_j^z\}$ for all $j$ in $\{1,2,\ldots,L\}$
\be
\!\!\mathcal B_L \!=\! \left\{\ket{\boldsymbol s} \!=\! \ket{s_1,\ldots,s_L}\!,\, s_j\!\in\!\{\pm1\}\!: \;\sigma_j^z\ket{\boldsymbol s} = s_j \ket{\boldsymbol s} \right\}\!.
\label{eq:computationalbasis}
\ee      
In this paper we interchangeably use ${s=+1\equiv\,\uparrow}$ and ${s=-1\equiv\,\downarrow}$ to designate eigenvalues of Pauli matrices. We stress that, as proven in Ref.~\cite{letter}, the self-dual kicked Ising model is ergodic for any $\boldsymbol h \neq 0$. This means that, due to its special structure, the model never displays Floquet-MBL~\cite{MBL,Papic, Roderich}. Note that the special structure of the self-dual model has recently been used to design a multiparty entanglement generation algorithm~\cite{NSM:multipartyentanglement}.

The Floquet operator generated by \eqref{eq:ham} reads as 
\be
U_{\rm KI}[\boldsymbol h]= 
{\frak T}\!\exp\!\!\left[-i\int_0^1\!\!\!\!{\rm d}t\, H_{\rm KI}[\boldsymbol h;t]\right]\!\!=
e^{-i H_{\rm K}}e^{-i H_{\rm I}[\boldsymbol h]}\,,
\label{eq:floquet}
\ee
where ${\frak T}\!\exp\left[\cdot\right]$ denotes a time-ordered exponential. The time-evlution generated by \eqref{eq:statet} admits a straightforward local 2-qubit quantum circuit representation. This is observed by writing  
\be
e^{-i H_{\rm I}[\boldsymbol h]}=\prod_{i\in\rm odd}e^{-i\frac{\pi}{4}  \sigma^{z}_j \sigma^z_{j+1}}\prod_{i\in\rm even}e^{-i\frac{\pi}{4}  \sigma^{z}_j \sigma^z_{j+1}}\,.
\ee

To drive the system out of equilibrium we consider a global quantum quench protocol: we initialise the system in the ground state of a short-range Hamiltonian and suddenly, say at $t=0$, we start evolving with \eqref{eq:ham}. In particular, here we consider the ground states $\ket{\psi_{\boldsymbol\theta,\boldsymbol\phi}}$ of the following family of local, non-interacting, magnetization Hamiltonians
\be
H_{\boldsymbol \theta,\boldsymbol \phi}\!= \!-\!\sum_{j=1}^{L} 
 g_j {\vec n}_{\theta_j,\phi_j}\cdot\vec{\sigma}_j,
\label{eq:magham}
\ee
where ${\boldsymbol\theta=(\theta_{1},\ldots,\theta_{L})}$ and ${\boldsymbol\phi=(\phi_{1},\ldots,\phi_{L})}$ are $L$-component vectors with components ${\theta_j \in [0,\pi]}$ and ${\phi_j \in [0,2\pi]}$, while $g_j>0$ is arbitrary, and
\be
\vec{ n}_{\theta,\phi} =\left( \sin\theta\cos\phi, \sin\theta\sin\phi,\cos\theta\right),
\ee
is the radial unit vector in the three-dimensional space.    

The ground states 
$\ket{\psi_{\boldsymbol\theta,\boldsymbol\phi}}$ of  (\ref{eq:magham}) are separable (\emph{i.e.} they have zero entanglement): the spin at site $j$ points in the direction $\vec { n}_{\theta_j,\phi_j}$ of its Bloch's sphere. Namely, the states $\ket{\psi_{\boldsymbol\theta,\boldsymbol\phi}}$ are explicitly written as   
\be
\label{eq:state0}
 \ket{\psi_{\boldsymbol\theta,\boldsymbol\phi}} =\bigotimes_{j=1}^{L}\left[\cos\!\left(\frac{\theta_j}{2}\right)\ket{\uparrow} +\sin\!\left(\frac{\theta_j}{2}\right) e^{i \phi_j}\ket{\downarrow}\right]\,. 
\ee
 After $t$ periods of the Floquet evolution the state of the system then reads 
\be
\ket{\psi_{\boldsymbol\theta,\boldsymbol\phi}(t)}=(U_{\rm KI}[\boldsymbol h])^t\ket{\psi_{\boldsymbol\theta,\boldsymbol\phi}}\,.
\label{eq:statet}
\ee
We stress that this protocol is constructive and simple enough to be realizable experimentally, for instance in the context of cold atoms~\cite{coldatoms,coldatoms1,coldatoms2}. 

\begin{figure}[b]
\includegraphics[width=\columnwidth]{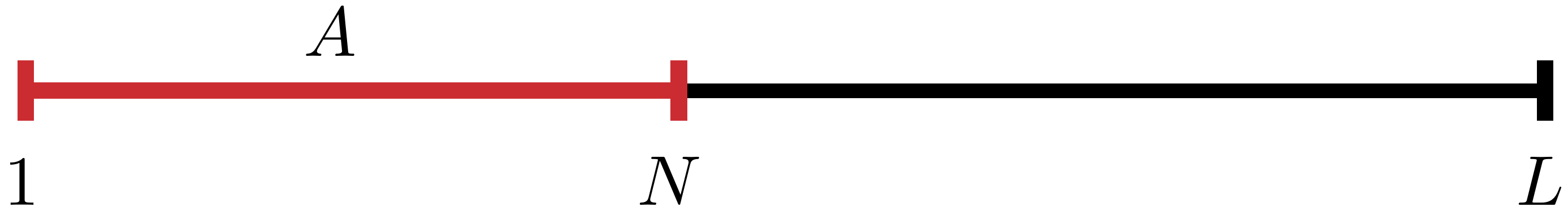}
\caption{The partition of the spin chain considered in the calculation of the entanglement.}
\label{fig:bipartition}
\end{figure}
In this work, the dynamics of the system are characterised by studying the time evolution of the entanglement between a contiguous subset of $N$ spins ${A=\{1,2,\ldots N\}}$ and the rest of the system, see Fig.~\ref{fig:bipartition}. This is encoded in the density matrix of the system reduced to the subsystem $A$, defined as   
\be
\rho_A(t)= {\rm tr}_{\mathcal H_{L-N}}\!\bigl[\ket{\psi_{\boldsymbol\theta,\boldsymbol\phi}(t)}\bra{\psi_{\boldsymbol\theta,\boldsymbol\phi}(t)}\bigr]\,,
\label{eq:reduceddensitymatrix}
\ee
where ${\cal H}_{L-N}$ is the Hilbert space associated with the complement $A^{\rm c} = \{N+1,\ldots L\}$ of $A$. The entanglement content of $\rho_A(t)$ is quantified by the R\'enyi entropies $S^{(\alpha)}_A(t)$, also called entanglement entropies. These are a one-parameter family of functionals of $\rho_A(t)$ defined as follows
\be
S^{(\alpha)}_A(t)=\frac{1}{1-\alpha}\log {\rm tr}\left[(\rho_A(t))^\alpha\right]\,,\qquad \alpha>0\,.
\label{eq:Renyi}
\ee
A particularly important member of this family is the von Neumann entropy 
\be
S^{(\rm vN)}_A(t) = \lim_{\alpha \to 1} S^{(\alpha)}_A(t) = - \log {\rm tr}\left[\rho_A(t)\log\rho_A(t)\right]\,,
\label{eq:vonNeumann}
\ee
which is the most used measure of bipartite entanglement for pure states~\cite{Entrev1}.

In summary: in this paper we study the time evolution generated by \eqref{eq:ham} of the R\'enyi entropies \eqref{eq:Renyi} for a system initialised in the states $\ket{\psi_{\boldsymbol\theta,\boldsymbol\phi}}$. As we shall see, our analytical results apply in the thermodynamic limit ${L\to \infty}$. We stress that, in contrast with what we did elsewhere~\cite{letter}, we do not introduce any averaging on the longitudinal magnetic fields. The thermodynamic limit is taken for fixed initial state and time-evolving Hamiltonian. 

%%%%%%%%%%%%%%%%%
\section{Outline of the results}%%
%%%%%%%%%%%%%%%%%
\label{sec:results} 

Our main result consists in finding two specific but physically interesting subclasses of the states \eqref{eq:state0} for which the time evolution of all R\'enyi entropies in the thermodynamic limit can be found exactly, for any configuration of magnetic fields $\{h_j\}$ and subsystem size $N$. These special classes of states are defined as 
\begin{align}
&\mathcal T = \left\{\ket{\psi_{{\frac{\pi}{2}}\boldsymbol 1,\boldsymbol \phi}},\quad \phi_j\in[0,2\pi]\right\},\label{eq:classesT}\\
&\mathcal L = \left\{\ket{\psi_{\boldsymbol{ \bar \theta}, \boldsymbol{\phi}}},\quad\quad\, \bar \theta_j\in\{0,\pi\}\right\},
\label{eq:classesL}
\end{align}
where $\boldsymbol 1$ denotes a vector of length $L$ with all entries equal to $1$~\cite{note3}. We respectively name them ``transverse separating states'' and ``longitudinal separating states'', while we generically call ``separating state'' a state belonging to either $\mathcal T$ or $\mathcal L$. These states are called ``transverse'' and ``longitudinal'' because they are respectively eigenstates of the operators $\cos\phi_j \sigma^x_j+\sin\phi_j \sigma^y_j$ and $\sigma^z_j$ for all $j$s. Therefore, they can be thought of as  configurations of a magnet where the spins lie on the $x$-$y$ plane (``transverse plane") or along the $z$ axis (``longitudinal axis"). The adjective ``separating'' refers to their key mathematical property and it is thoroughly  explained in Sec.~\ref{sec:separating}. We stress that the property of being separating is more restrictive that being just separable: all the states $\ket{\psi_{\boldsymbol{\theta}, \boldsymbol{\phi}}}$ are separable but only a subset of them are separating. Specific instances of separating states, which are most relevant from the experimental point of view, are the ground states of the two parts in the Floquet protocol. For example, when $J>0$ and $|h_j|<J$ the ground state of $H_{\rm I}$ is $\ket{\psi_{\pi\boldsymbol 1,\boldsymbol 0}}\in{\cal L}$, while when $b>0$, the ground state of $H_{\rm K}$ is $\ket{\psi_{\frac{\pi}{2}\boldsymbol 1,\pi\boldsymbol 1}}\in{\cal T}$.

To simplify the analysis of the results it is useful to note that the time evolution of the states in ${\cal L}$ can be related to that of those in $\mathcal T$. This is easily seen by means of the following identity
\be
\ket{\psi_{\bar{\boldsymbol\theta},\boldsymbol\phi}(1)}=U_{\rm KI}[\boldsymbol h]\ket{\psi_{\bar{\boldsymbol\theta},\boldsymbol\phi}}
\simeq \ket{\psi_{{{\frac{\pi}{2}}\boldsymbol 1,\boldsymbol{ \bar \theta}-{\frac{\pi}{2}}\boldsymbol 1}}},
\label{eq:identitystates}
\ee
where $\simeq$ denotes equality up to a global phase. This identity is proven by observing that, since the states in $\cal L$ are eigenstates of $\sigma^z_j$, they are also eigenstates of $H_{\rm I}[\boldsymbol h]$. Therefore the application of $e^{-i H_{\rm I}[\boldsymbol h]}$ only changes $\ket{\psi_{\bar{\boldsymbol\theta},\boldsymbol\phi}}$ by a global phase. Moreover, an explicit calculation shows that
\be 
e^{-i H_{\rm K}}\ket{\psi_{\bar{\boldsymbol\theta},\boldsymbol\phi}}\simeq \ket{\psi_{{{\frac{\pi}{2}}\boldsymbol 1,\boldsymbol{ \bar \theta}-{\frac{\pi}{2}}\boldsymbol 1}}}.
\ee
Eq.~\eqref{eq:identitystates} means that the first time step of evolution for states in $\cal L$ keeps them in a separable form, and hence does not change the entanglement, but turns them into states in $\cal T$. An immediate consequence of Eq.~\eqref{eq:identitystates} is
\be
\ket{\psi_{\bar{\boldsymbol\theta},\boldsymbol\phi}(t)}\simeq \ket{\psi_{\frac{\pi}{2}\boldsymbol{1},\boldsymbol{ \bar \theta}-{\frac{\pi}{2}}\boldsymbol 1}(t-1)},\qquad t\geq1\,.
\label{eq:TLmapping}
\ee
Considering the entanglement entropy we then have
\be
S^{(\alpha)}_A(t)|_{\bar{\boldsymbol\theta},\boldsymbol\phi}=S^{(\alpha)}_A(\max(t-1,0))|_{\frac{\pi}{2}\boldsymbol{1},\boldsymbol{ \bar \theta}-{\frac{\pi}{2}}\boldsymbol 1}\,,\,\,\, t\geq0\,,
\label{eq:LtoT}
\ee
where we explicitly reported the initial state dependence and used $S^{(\alpha)}_A(0)=0$. By virtue of this simple argument we can restrict our attention to the states in $\cal T$, the time evolution of the entropy for the states in $\cal L$ is found using Eq.~\eqref{eq:LtoT}.  

\begin{figure}[t]
\includegraphics[width=0.48\textwidth]{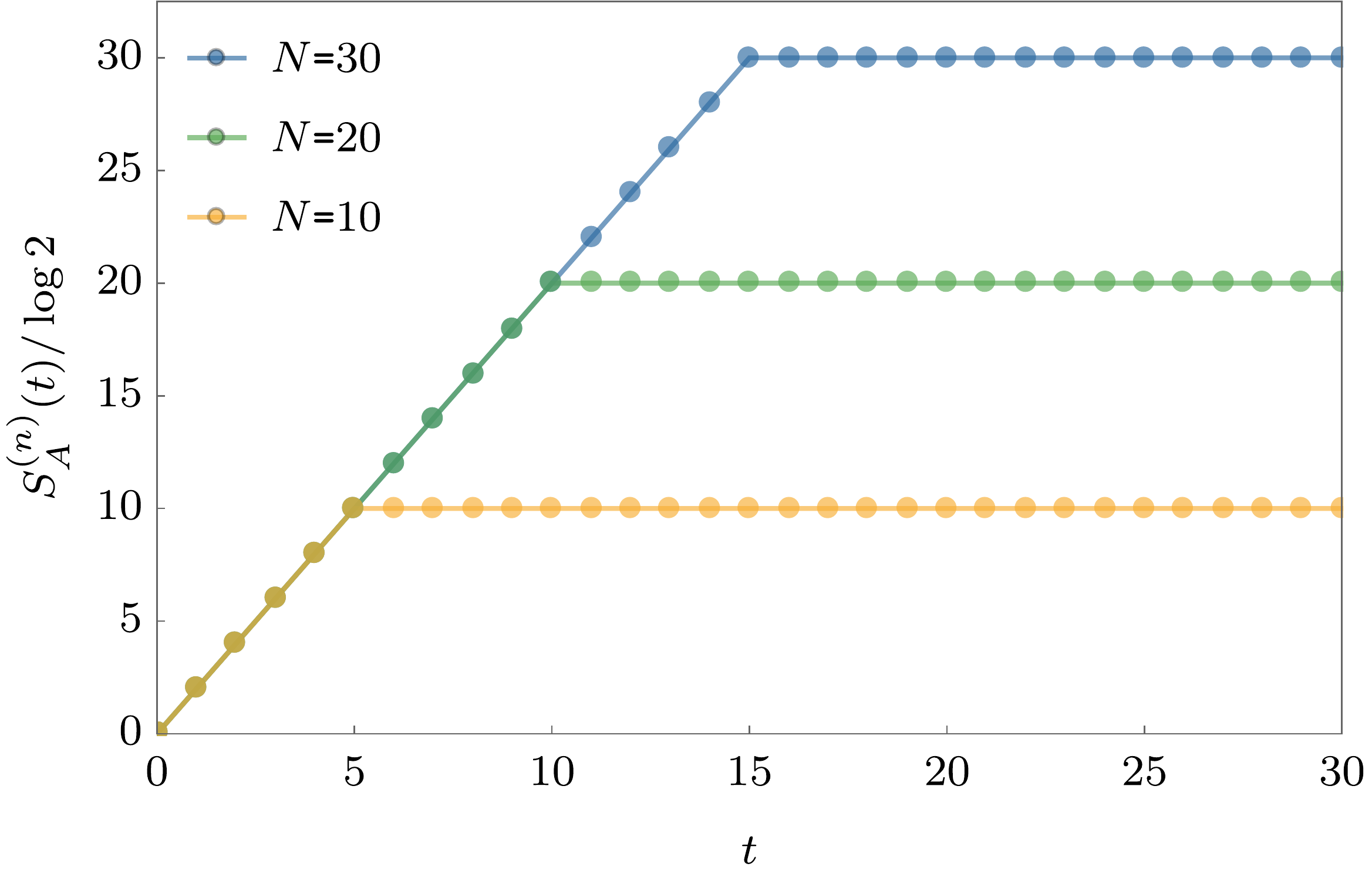}
\caption{Plot of the exact result of all R\'enyi entropies given by \eqref{eq:finalresult} for $\theta=\pi/2$ case. The expressions for $\theta=0$ are delayed by one period.}
\label{fig:plotresult}
\end{figure}

The time evolution of entanglement entropies from states in ${\cal T}$ (in the thermodynamic limit) can be explicitly determined by means of the ``duality method'' developed in Secs.~\ref{sec:duality}--\ref{sec:specialcase}. The result reads as  
\be
\lim_{L\rightarrow \infty} S^{(\alpha)}_A(t)= \min(2t,N) \log 2\,.
\label{eq:finalresult}
\ee
This result is indisputably remarkable: when evolving from separating states all entanglement entropies take the same universal form, independent of the fields $h_j$ and of details of the initial states. In particular, at fixed $N$, the entanglement entropies display a linear growth in time up to a non-analytic saturation point where they become constant, see Fig.~\ref{fig:plotresult}.
The independence of $n$ of the result means that the spectra of the reduced density matrices $\rho_A(t)$ are {\em flat}. In other words the reduced density matrices have exactly ${2^{\min(2t,N)}}$ eigenvalues equal to ${2^{-\min(2t,N)}}$ while the others vanish. 

A form like \eqref{eq:finalresult} for the evolution of the entanglement entropies has been found in a number of different physical settings, both in closed and periodically driven systems. Examples range from conformal invariant systems \cite{CC, LS:CFT}, to non-integrable closed systems \cite{KH:NonIntEnt}, from random in time~\cite{Nahum2, r-2017, RandomCircuitsEnt}, to periodically driven~\cite{CDC1} random unitary circuits. As opposed to all these cases, however, our result does not hold only at the leading order for large $t$ and $N$. It holds with no corrections for any $t$ and $N$. This gives further evidence for the special status of the self-dual kicked Ising model as minimal solvable model for quantum chaos.

Another interesting feature of our result is that is saturates the ``minimal cut" bound~\cite{CLM:minimalcut}. The bound states that, when evolving from a product state, 
\be
S^{(\alpha)}_A(t)\leq \ell_{\rm min} \log q\,,\qquad \forall \alpha\,,
\label{eq:mincut}
\ee
where $q$ is the dimension of the local Hilbert space (2 in our case) and $\ell_{\rm min}$ is the minimal number of links intersected by a cut separating the region A from the rest of the system in a local quantum circuit representation of 
$U_{\rm KI}^t$. The fact that the bound is saturated means that entanglement spreads with the maximal possible speed allowed by the range of the Hamiltonian and the dimension of the local Hilbert space. This can be seen in a more physical way by noting that \eqref{eq:finalresult} implies that, for $t\leq N/2$, at each time step two more spins of the block $A$ become maximally entangled with the rest of the system. Following Ref.~\cite{LS:CFT}, this can be imagined as an entanglement wave propagating into the block $A$ from the two boundaries. The fact that our exact result \eqref{eq:finalresult} saturates the bound \eqref{eq:mincut} also means that it agrees with the ``minimal membrane'' picture recently put forward in Ref.~\cite{RandomCircuitsEnt}, where a coarse grained version of the minimal cut has been proposed to describe the leading-order-in-time features of the entanglement dynamics in generic systems. Interestingly, however, our system also contains an integrable point, namely $\boldsymbol{h}=0$. At this point our result agrees with the the quasiparticle picture of Ref.~\cite{CC}, because in our case all quasiparticles move at the same, maximal, speed ($|v|=1$). We note that, at the integrable point, the result \eqref{eq:finalresult} has also been found in Ref.~\cite{MLS:multipartiteentanglement}, for the evolution of the von Neumann entanglement from the separating state $\ket{\psi_{{\frac{\pi}{2}}\boldsymbol 1,\boldsymbol 0}}$.

If the initial state is not separating the problem is not amenable to an analytical treatment. In Sec.~\ref{sec:genericcase}, however, performing a thorough numerical analysis we argue that the entanglement spreading from all the states~\eqref{eq:state0} is still described by \eqref{eq:finalresult} at the leading order in time, provided that the system is away from the integrable point $\boldsymbol h=\boldsymbol 0$. 
Note that this conjecture is physically very reasonable: since the system is quantum ergodic it is natural to expect the entanglement entropies to eventually become state-independent. On the contrary, in the integrable case our numerical results are consistent with 
\be
\lim_{L\rightarrow \infty} S^{(\alpha)}_A(t)= \min(2t,N)S^{(\alpha)}_{\boldsymbol \theta,\boldsymbol \phi}\,,
\label{eq:finalresultint}
\ee
where $S^{(\alpha)}_{\boldsymbol \theta,\boldsymbol \phi}\leq \log 2$ is an initial-state-dependent constant. In the numerical analysis that led to these results, a crucial role has been played by a duality-based numerical approach (see Sec.~\ref{sec:genericcase}) that allows us to treat the system in the infinite volume limit. Supplemented with some analytical information it can reach $t=17$ Floquet periods of evolution even in cases where the entanglement grows at the maximal speed. Even if based on the ``duality-method", this approach does not rely on the special algebraic structure arising at the self dual points \eqref{eq:selfduals} and it is applicable in the whole parameter space of the kicked Ising model. 

Another marked difference between the integrable and non-integrable case is observed when the system is confined in a finite volume $L$. Indeed, in this setting the integrable system displays finite-size related recurrences when $t\sim L$, while these recurrences are absent (or at least negligible) in the non-integrable case. These results respectively agree with the predictions of the quasiparticle and the minimal membrane pictures and are also consistent with the numerical analysis of the entanglement spreading in the kicked Ising model carried out in Ref.~\cite{PL:kickedIsing}.

%%%%%%%%%%%%%%%%%%%%%%%%%%%%%%%%%%%%%%%%%%%%%%%%%%%%%%%%%%
\section{Duality Mapping for the Entanglement Entropies}%%
%%%%%%%%%%%%%%%%%%%%%%%%%%%%%%%%%%%%%%%%%%%%%%%%%%%%%%%%%%
\label{sec:duality}

\begin{figure}[t]
\includegraphics[width=\columnwidth]{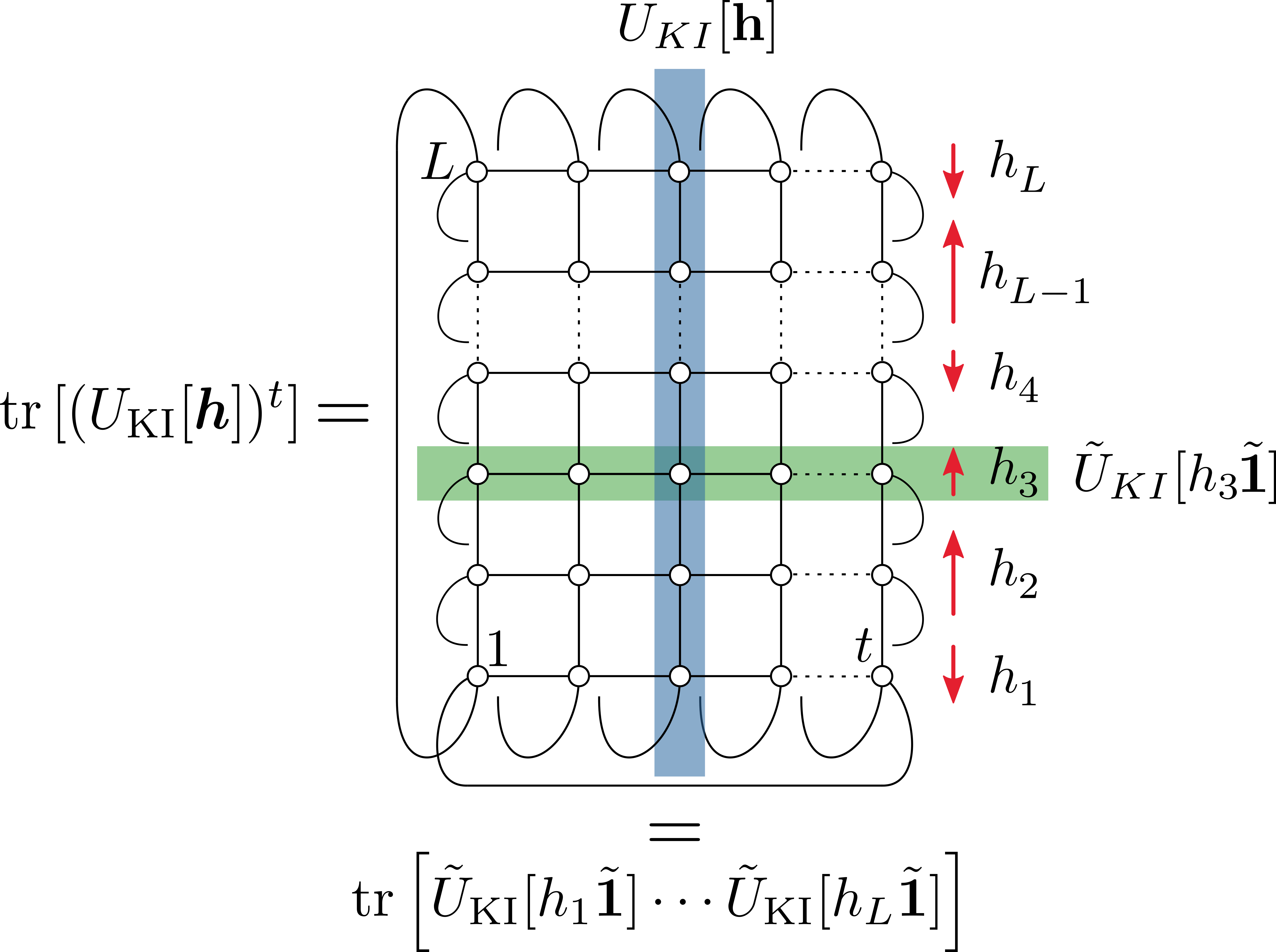}
\caption{Pictorial representation of the duality relation \eqref{eq:dual} fulfilled by the Floquet operator \eqref{eq:floquet}. Traces of powers of the Floquet operator correspond to the partition function of a classical Ising model on a $t\times L$ lattice. The column-to-column transfer matrix is given by $U_{\rm KI}[\boldsymbol h]$ while the row-to-row transfer matrix between the $(j-1)$-th and the $j$-th row is given by $\tilde U_{\rm KI}[h_j \tilde{\boldsymbol 1}]$. Note that self-duality condition implies that both transfer matrices are unitary.}
\label{fig:duality}
\end{figure}

In Ref.~\cite{Guhr} the authors pointed out that the traces of integer powers of the Floquet operator \eqref{eq:floquet} enjoy a useful space-time ``duality symmetry'', which can be demonstrated as follows. First note that 
\be
{\rm tr}\left[(U_{\rm KI}[\boldsymbol h])^t\right],\qquad\qquad t\in\mathbb N\,,
\ee
can be represented as a partition function of a classical Ising model on a $t\times L$ lattice, where $U_{\rm KI}[\boldsymbol h]$ acts as transfer matrix in time, see Fig.~\ref{fig:duality} for a pictorial representation and Appendix~\ref{app:duality} for the explicit expression. 
Second, observe that, due to the short-range nature of the couplings in \eqref{eq:hamiltonians1} and \eqref{eq:hamiltonians2}, the same quantity can also be written in terms of a transfer matrix defined on a lattice of $t$ sites and propagating in space (see Appendix~\ref{app:duality}). Namely we have 
\be
{\rm tr}\left[(U_{\rm KI}[\boldsymbol h])^t\right] = {\rm tr} \left[\tilde{U}_{\rm KI}[h_1 \tilde{\boldsymbol 1}]\cdots\tilde{U}_{\rm KI}[h_L \tilde{\boldsymbol 1}]\right],
\label{eq:dual}
\ee
where ``tilded''  bold symbols denote vectors of $t$ components, in particular 
\be
\tilde{\boldsymbol 1}=(\underbrace{1,\ldots,1}_{t}),
\ee
has all entries equal to $1$, and $\tilde{U}_{\rm KI}[\tilde{\boldsymbol h}]$ is the transfer matrix in space, also called ``dual'' transfer matrix. It turns out that $\tilde{U}_{\rm KI}[\tilde{\boldsymbol h}]$ has  the same form as the Floquet operator \eqref{eq:floquet} (with $L$ replaced by $t$ in \eqref{eq:hamiltonians1} and \eqref{eq:hamiltonians2}) where the longitudinal magnetic field vector is given by $\tilde{\boldsymbol h}$, while Ising coupling $\tilde J$ and the transverse field $\tilde b$ are given by the following functions of $J$ and $b$
\begin{align}
\tilde J &= -\frac{\pi}{4}-\frac{i}{2}\log\tan b\,, \label{eq:tildeJ}\\
\tilde b &= -\frac{\pi}{4}-\frac{i}{2}\log\tan J\,. \label{eq:tildeb}
\end{align}
Since $\tilde J$ and $\tilde b$ are generically complex, the transfer matrix $\tilde{U}_{\rm KI}[ \tilde{\boldsymbol h}]$ is generically not unitary. The dual couplings become real only when the model is at one of the self-dual points \eqref{eq:selfduals}.   

In Ref.~\cite{letter} we showed that such duality symmetry can be used to compute non-trivial observables, considering the example of the disorder averaged spectral form factor. In that case, even if the quantity cannot be written in terms of a transfer matrix in time, it can still be written in terms of a transfer matrix in space. This allowed us to perform an analytical calculation. The unitarity of the matrix $\tilde{U}_{\rm KI}[\tilde{\boldsymbol h}]$, however, proved itself to be a necessary requirement for the analytical approach to be feasible. This clarifies the special status of the self dual points \eqref{eq:selfduals}: they are the only points of the parameter space where this duality mapping leads to an analytic solution. 

Here we develop a similar duality mapping for the calculation of the entanglement entropies, or, more precisely, of the traces of integer powers of the reduced density matrix $\rho_A(t)$. We will see that also ${\rm tr}\left[(\rho_A(t))^n\right]$ can be written as the trace of a power of an appropriate transfer matrix in time. In Secs.~\ref{sec:separating} and~\ref{sec:specialcase} we then show that, at the self dual points and for the special initial states \eqref{eq:classesT} and \eqref{eq:classesL}, such a trace can be analytically evaluated.

Considering ${\rm tr}\left[(\rho_A(t))^n\right]$ and using the definitions \eqref{eq:statet} and \eqref{eq:reduceddensitymatrix} we find   
\begin{align}
&\!\!{\rm tr}\left[(\rho_A(t))^n\right] = \notag\\
&\!\!\!\sum_{\{ \boldsymbol a_{i}\},\{\boldsymbol b_{i}\}}\!\!\!\!\!\!\bra{\psi_{\boldsymbol\theta,\boldsymbol\phi}}(U_{\rm KI}[\boldsymbol h])^{-t}\!\ket{\boldsymbol a_1,\boldsymbol b_2}\! \bra{\boldsymbol a_1, \boldsymbol b_1}(U_{\rm KI}[\boldsymbol h])^{t}\!\ket{\psi_{\boldsymbol\theta,\boldsymbol\phi}}\notag\\
&\times\!\!\bra{\psi_{\boldsymbol\theta,\boldsymbol\phi}}(U_{\rm KI}[\boldsymbol h])^{-t}\!\ket{\boldsymbol a_2,\boldsymbol b_3}\! \bra{\boldsymbol a_2, \boldsymbol b_2}(U_{\rm KI}[\boldsymbol h])^{t}\!\ket{\psi_{\boldsymbol\theta,\boldsymbol\phi}}\notag\\
&\qquad\qquad\qquad\qquad\quad\qquad\;\,\vdots\notag\\
&\times\!\!\bra{\psi_{\boldsymbol\theta,\boldsymbol\phi}}(U_{\rm KI}[\boldsymbol h])^{-t}\!\ket{\boldsymbol a_n,\boldsymbol b_1}\! \bra{\boldsymbol a_n, \boldsymbol b_n}(U_{\rm KI}[\boldsymbol h])^{t}\!\ket{\psi_{\boldsymbol\theta,\boldsymbol\phi}}
\label{eq:trrhoA}
\end{align}
where 
$\ket{\boldsymbol a_i,\boldsymbol b_j}=\ket{\boldsymbol a_i}\otimes\ket{\boldsymbol b_j}$, $\ket{\boldsymbol a_i}\in \mathcal B_{N}$, $\ket{\boldsymbol b_i}\in \mathcal B_{L-N}$, for $i,j\in\{1,2,\ldots,n\}$. Here we denoted by $\mathcal  B_j$ the computational basis of 
\be
{\cal H}_j=(\mathbb C^2)^{\otimes j}\,.
\ee
An explicit expression of ${\cal B}_{j}$ is obtained replacing $L$ by $j$ in the expression \eqref{eq:computationalbasis}.

\begin{figure*}[t]
\includegraphics[width=0.75\textwidth]{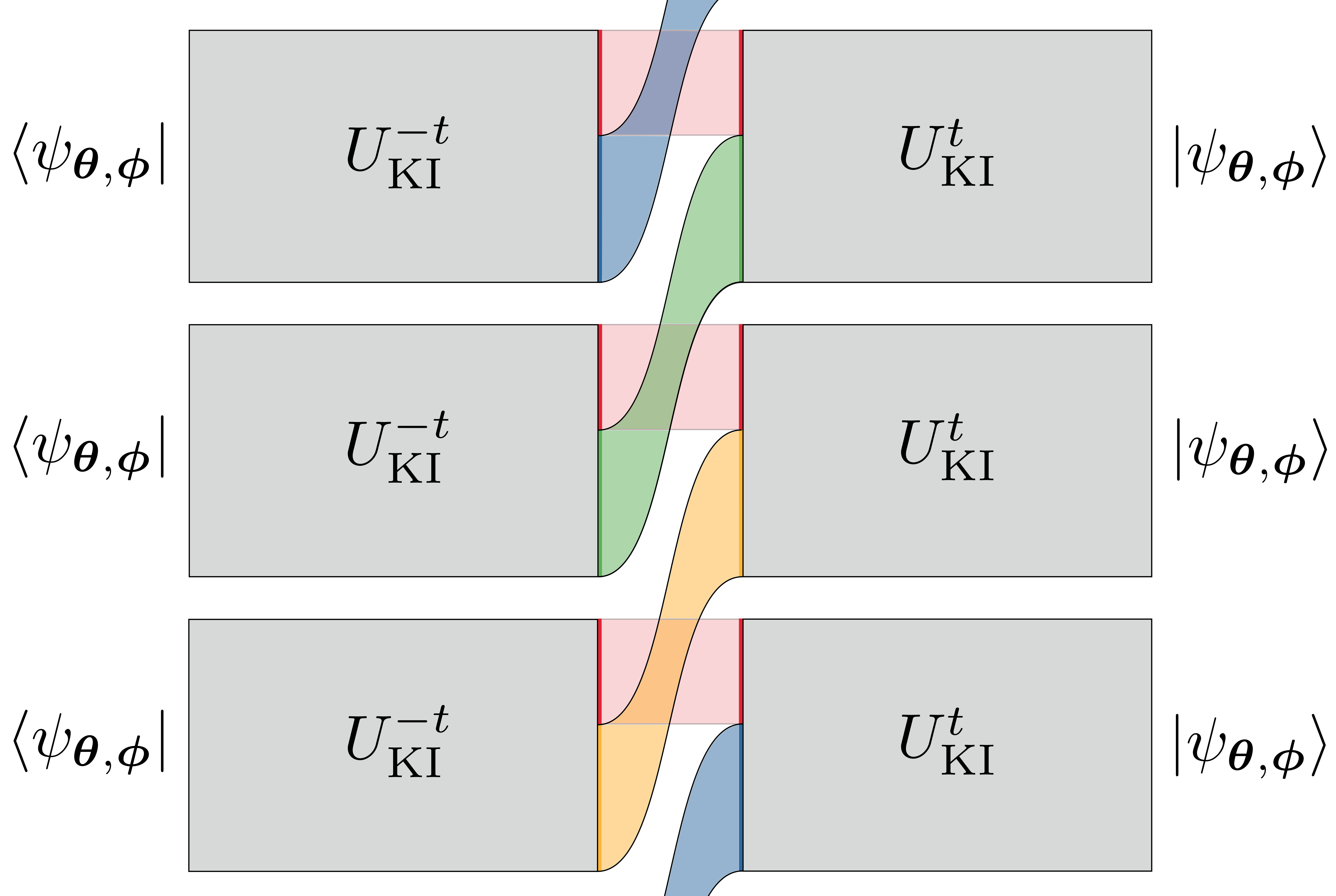}
\caption{Schematic representation of ${\rm tr}\left[(\rho_A(t))^3\right] $ according to the expression (\ref{eq:trrhoA}). The six different cylinders corresponding to the partition functions \eqref{eq:singlecylinder} are schematically represented as rectangles. The spin subchains $A$ and $A^{\rm c}$ connected with the coloured belts share identical spin configurations.}
\label{fig:powersrenyi}
\end{figure*}

Eq.~\eqref{eq:trrhoA} allows one to interpret the trace of the $n$-th power of the reduced density matrix as the partition function of a classical statistical mechanical model on a multi-sheeted two-dimensional lattice, see Fig.~\ref{fig:powersrenyi} for a pictorial representation in the case $n=3$. To see it more explicitly we consider a single ``building block"
\be
\bra{\boldsymbol a,\boldsymbol b}(U_{\rm KI}[\boldsymbol h])^{t}\!\ket{\psi_{\boldsymbol\theta,\boldsymbol\phi}}\,,
\label{eq:BB}
\ee
and show that it is equivalent to the partition function of a classical Ising model (with complex weights) on a ${t \times L}$ lattice with periodic boundary conditions in space and fixed boundary conditions in time.  
This is seen in two steps. First, we insert $t$ resolutions of the identity operator in the basis \eqref{eq:computationalbasis} into \eqref{eq:BB}, obtaining 
\begin{align}
\!\!\bra{\boldsymbol a,\boldsymbol b}(U_{\rm KI}[\boldsymbol h])^{t}\ket{\psi_{\boldsymbol\theta,\boldsymbol\phi}}\!=&\!\!  \sum_{\{\boldsymbol s_{\tau}\}} \prod_{\tau=1}^{t-1} \!\bra{\boldsymbol s_{\tau+1}}\!U_{\rm KI}[\boldsymbol h]\!\ket{\boldsymbol s_{\tau}}\notag\\
&\,\times\!\!\bra{\boldsymbol a,\boldsymbol b}\!U_{\rm KI}[\boldsymbol h]\!\ket{\boldsymbol s_{t}}\!\braket{\boldsymbol s_{1}|\psi_{\boldsymbol\theta,\boldsymbol\phi}}\!.
\label{eq:resolutionId}
\end{align}
Then we evaluate the matrix elements 
\be
\!\!\braket{\boldsymbol s|\psi_{\boldsymbol\theta,\boldsymbol\phi}}\!=\! \prod_{j=1}^L\!\!\left[ \cos(\theta_j/2)\delta_{s_{j},1}\!+\!\sin(\theta_j/2) e^{i\phi_j} \delta_{s_{j},-1}\right]\!,
\ee
and 
\begin{align}
&\bra{\boldsymbol s}\!U_{\rm KI}[\boldsymbol h]\!\ket{\boldsymbol r}\!\!=\!\! \left(\frac{i}{2}\right)^{\!\!\frac{L}{2}} \!\!\!\!\exp\!\Biggl[
- {\frac{i\pi}{4} \!\sum_{j=1}^L s_{j} r_{j}}\Biggr]\notag\\
&\quad\quad\qquad\qquad\times\!\exp\!\Biggl[- {\frac{i\pi}{4}\! \sum_{j=1}^L r_{j} r_{j+1}- i \!\sum_{j=1}^L h_j r_{j}}\Biggr]\!,
\label{eq:monstereq}
\end{align}
Here, to find the last equation we set $r_{L+1}=r_1$ and we used the identity
\be
\braket{s|e^{i \frac{\pi}{4} \sigma^x}|r}= \sqrt{\frac{i}{2}} \exp\left[ - i\frac{\pi}{4}  s r\right],\qquad s,r\in\{\pm1\}\,,
\ee 
to treat the ``kick'' part of the Floquet operator $U_{\rm KI}[\boldsymbol h]$. Putting all together we have 
\begin{widetext}
\begin{align}
\bra{\boldsymbol a,\boldsymbol b}\!(U_{\rm KI}[\boldsymbol h])^{t}\!\ket{\psi_{\boldsymbol\theta,\boldsymbol\phi}}\!=\!& \left(\frac{i}{2}\right)^{\!\!\frac{t L}{2}} \!\!\!\!  \sum_{\{ s_{\tau,j}\}}\! \exp\!\Biggl[-\frac{i\pi}{4} \sum_{\tau=1}^{t} \sum_{j=1}^L s_{\tau,j} s_{\tau,j+1}- \frac{i\pi}{4} \sum_{\tau=1}^{t-1}\sum_{j=1}^Ls_{\tau,j} s_{\tau+1,j}\! - i \!\sum_{\tau=1}^{t} \sum_{j=1}^L h_j s_{\tau, j}\Biggr]\notag\\
&\qquad\times\!\exp\!\Biggl[- \frac{i\pi}{4}\!\sum_{j=1}^N s_{t,j} a_{j}\!-\!\frac{i\pi}{4}\!\!\!\sum_{j=N+1}^L \!\!\!\!s_{t,j} b_{j-N}\!\Biggr]\prod_{j=1}^L\!\!\left[ \cos(\theta_j/2)\delta_{s_{1,j},1}+\sin(\theta_j/2) e^{i\phi_j} \delta_{s_{1,j},-1}\right]\!.
\label{eq:singlecylinder}
\end{align}  
which, as promised, is the partition function of the classical Ising model on a two-dimensional cylinder. 

Representing in this way each of the $2n$ building blocks in \eqref{eq:trrhoA} and summing over $\{\boldsymbol a_j,\boldsymbol b_j\}$, one connects together the $2n$ different cylinders obtaining the announced multi-sheeted lattice. Explicitly we have    
\begin{align}
{\rm tr}\left[(\rho_A(t))^n\right]=& \frac{1}{2^{n L t}}\!\!\! \sum_{\{ s_{\nu,\tau,j}\}} \!\!\!\exp\!\!\left[
-i \sum_{j=1}^L\sum_{\nu=1}^{2n} {\rm sgn}(n-\nu) \left( \sum_{\tau=1}^{t} \left( \frac{\pi}{4} s_{\nu,\tau,j} s_{\nu,\tau,j+1} +  h_j s_{\nu,\tau,j} \right) + \sum_{\tau=1}^{t-1} \frac{\pi}{4}s_{\nu,\tau,j} s_{\nu,\tau+1,j}\right)\right]\notag\\
& \qquad\qquad\times\prod_{\nu=1}^n\!\left\{\prod_{j=1}^{N} \!\!\left(1+s_{\nu,t,j}s_{\nu+n,t,j}\right)\!\!\!\!\!\prod_{j=N+1}^{L} \!\!\!\!\!\!\left(1+s_{\nu,t,j} s_{n+1+{\rm mod}(\nu-2,n),t,j}\right)\!\!\right\}\notag\\
& \qquad\qquad\times\prod_{\nu=1}^{2n} \prod_{j=1}^{L} \left(\cos(\theta_j/2) \delta_{s_{\nu,1,j},1}+\sin(\theta_j/2) e^{i\phi_j {\rm sgn}(n-\nu) } \delta_{s_{\nu,1,j},-1}\right)\,,
\label{eq:trrhoAexplicit}
\end{align}
\end{widetext}
where ${\rm sgn}(x)$ is the sign function (we adopted the convention ${{\rm sgn}(0)=1}$), ${{\rm mod}(m,n)=m\,{\rm mod}\,n}$ is the mod-function, and we introduced a new index ${\nu\in\{1,2,\ldots,2n\}}$ such that strings ${\boldsymbol s_{\nu,\tau}}$ with $\nu\le n$ belong to terms in \eqref{eq:trrhoA} with forward time evolution, while those with $\nu > n$ belong to terms in \eqref{eq:trrhoA} with backward time evolution. 
 
The second line of \eqref{eq:trrhoAexplicit} is obtained by explicitly summing over $\{ \boldsymbol a_{i}\}$ and $\{\boldsymbol b_{i}\}$ with the help of the identity 
\be
\!\!\!\sum_{a\in\{\pm 1\}}\exp\left[ - i\frac{\pi}{4}  a (s-r)\right]=1+s r,\quad\,\, s,r\in\{\pm1\}. 
\ee
We see that this line forces the configurations of spins in the subchains $A$ and $A^c$ on the edges of different cylinders to be the same. These ``frozen'' configurations are represented by coloured strips in Fig.~\ref{fig:powersrenyi}.  
 
To proceed, it is useful to introduce the tensor product space $\mathcal H_{t}^{\otimes 2n}$, composed of $2n$ copies of $\mathcal H_t$, which is the space where the dual Floquet operator $\tilde{U}_{\rm KI}[\tilde{\boldsymbol h}]$ acts. More formally
\be
 \mathcal H_{t}^{\otimes 2n}=\overbrace{\mathcal H_{t}\otimes\ldots\otimes\mathcal H_{t}}^{2n}\cong\mathcal H_{2 n t}\,.
\ee
Then, we define the operators $\mathbb T_{\theta,\phi}[h]$ and $\mathbb R_{\theta,\phi}[h]$ on $\mathcal H_{t}^{\otimes 2n}$ through their matrix elements in the computational basis        
\begin{widetext}
\begin{align}
\!\!\!\braket{\{s_{\nu,\tau}\}| \mathbb T_{\theta,\phi}[h]|\{r_{\nu,\tau}\}}=&\frac{1}{2^{(t-1) n}}
\exp\!\!\left[-i \sum_{\nu=1}^{2n} {\rm sgn}(n-\nu) \left( \sum_{\tau=1}^{t} \left( \frac{\pi}{4} s_{\nu,\tau} r_{\nu,\tau} +  h_j s_{\nu,\tau} \right) + \sum_{\tau=1}^{t-1} \frac{\pi}{4}s_{\nu,\tau} s_{\nu,\tau+1}\right)\right]\notag\\
&\times \prod_{\nu=1}^n \left(\frac{1+s_{\nu,t} s_{\nu+n,t}}{2}\right) \prod_{\nu=1}^{2n} \left( \cos(\theta/{2})\delta_{s_{\nu,1},1}+\sin({\theta}/{2})e^{i\phi\,{\rm sgn}(n-\nu)} \delta_{s_{\nu,1},-1}\right)\!,
\label{eq:matrixelementsT}
\end{align}
and 
\begin{align}
\!\!\!\!\braket{\{s_{\nu,\tau}\}| \mathbb R_{\theta,\phi}[h]|\{r_{\nu,\tau}\}}=&\frac{1}{2^{(t-1) n}} 
\exp\!\!\left[-i \sum_{\nu=1}^{2n} {\rm sgn}(n-\nu) \left( \sum_{\tau=1}^{t} \left( \frac{\pi}{4} s_{\nu,\tau} r_{\nu,\tau} +  h_j s_{\nu,\tau} \right) + \sum_{\tau=1}^{t-1} \frac{\pi}{4}s_{\nu,\tau} s_{\nu,\tau+1}\right)\right]\notag\\
&\times \prod_{\nu=1}^n \left(\frac{1+s_{\nu,t} s_{n+1+{\rm mod}(\nu-2,n),t}}{2}\right)\prod_{\nu=1}^{2n} \left( \cos(\theta/{2})\delta_{s_{\nu,1},1}+\sin({\theta}/{2})e^{i\phi\,{\rm sgn}(n-\nu)} \delta_{s_{\nu,1},-1}\right)\!,
\label{eq:matrixelementsR}
\end{align}
\end{widetext}
where the first subscript labels spin variables in the different copies of $\mathcal H_t$ composing $\mathcal H_{t}^{\otimes 2n}$. 

\begin{figure*}[t]
\includegraphics[width=0.75\textwidth]{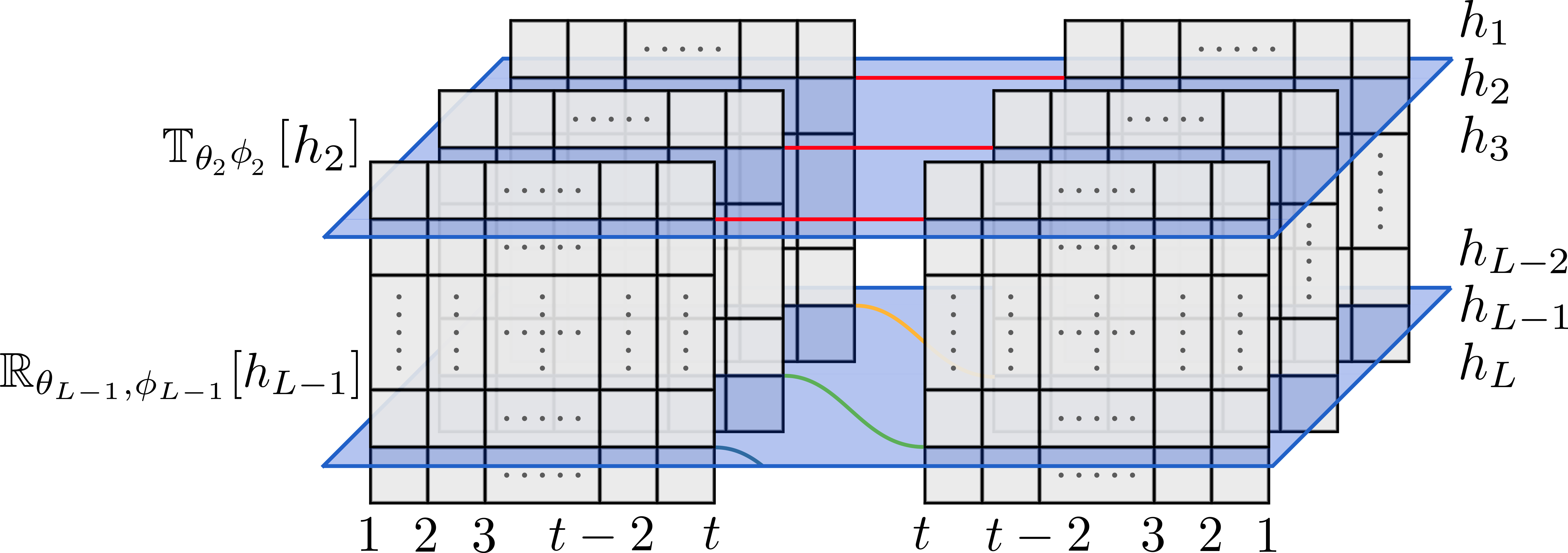}
\caption{Schematic depiction of  ${\rm tr}\left[(\rho_A(t))^3\right]$ written according to Eq.~\eqref{eq:traceintermediate}. Positive and negative time sheets are respectively on the left and on the right. Vertices connected by the coloured lines are coupled by the transfer matrices, in analogy with Fig.~\ref{fig:powersrenyi}. Blue-shaded horizontal planes denote the spatial transfer matrices, specifically the operator $\mathbb T_{\theta,\phi}[h]$
for the physical sites corresponding to the block $A$ of $N$ spins, and the operator 
$\mathbb R_{\theta,\phi}[h]=\mathbb P \mathbb T_{\theta,\phi}[h] \mathbb P^\dagger $ 
for the sites corresponding to $L-N$ spins in $A^{\rm c}$.}
\label{fig:dualityrenyi}
\end{figure*}
Using the above matrix elements it is immediate to see that the expression \eqref{eq:trrhoAexplicit} can directly be rewritten as a trace (on $\mathcal H_{t}^{\otimes 2n}$) of products of $\mathbb T_{\theta,\phi}[h]$ and $\mathbb R_{\theta,\phi}[h]$, namely 
\be
\!\!{\rm tr}\left[(\rho_A(t))^n\right]\!=\!{\rm tr}\!\!\left[\!\!\left(\prod_{j=1}^N \mathbb T_{\theta_j,\phi_j}[h_j]\!\!\right)\!\!\! \left (\prod_{j=N+1}^L \!\!\!\!\mathbb R_{\theta_{j},\phi_{j}}[h_{j}]\!\!\right)\!\!\right]\!\!,
\label{eq:traceintermediate}
\ee
where we defined on {\em ordered} product of non-commuting operators $\{\mathbb O_{j}\}$ as
\be
\prod_{j=a}^b \mathbb O_{j} =
\begin{cases}
\mathbb O_{a}\cdots \mathbb O_{b} &\text{if}\quad a\leq b\\
\1 &\text{if}\quad a>b
\end{cases}\,.
\ee
The rewriting achieved by \eqref{eq:traceintermediate} is pictorially represented in Fig.~\ref{fig:dualityrenyi}, again in the case $n=3$.

\begin{figure}[t]
\includegraphics[width=0.45\textwidth]{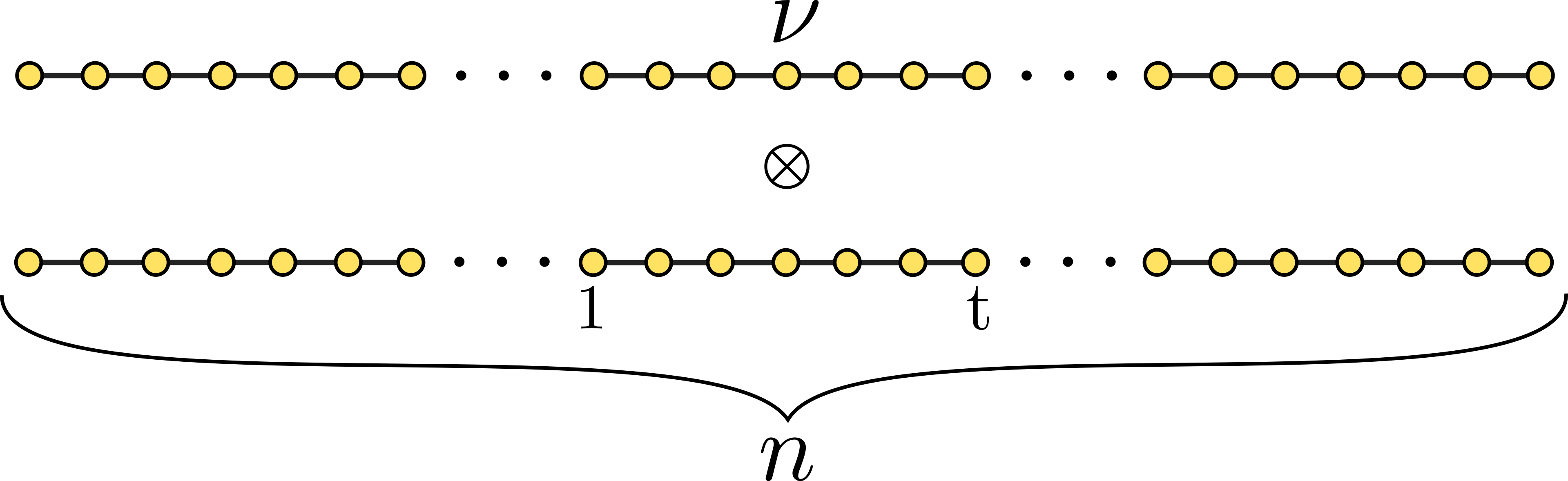}
\caption{Pictorial representation of the arrangement of the ``dual'' quantum spin degrees of freedom adopted in the tensor product space $\mathcal H_{t}^{\otimes 2n}$.}
\label{fig:tensorproductspace}
\end{figure}

In upcoming analysis it will be useful to think of $\mathcal H_{t}^{\otimes 2n}$ as a tensor product of two copies of $\mathcal H_{nt}$, grouping together the first and the last $n$ copies of $\mathcal H_t$, see Fig.~\ref{fig:tensorproductspace}. Namely we write each element of the basis of $\mathcal H_{t}^{\otimes 2n}$ in the following way 
\be
\ket{\{s_{a,\tau}\}^{1\le a\le 2n}_{1\le \tau\le t}} =
\ket{\{s_{a,\tau}\}^{1\le a\le n}_{1\le \tau\le t}}
\otimes 
\ket{\{s_{a,\tau}\}^{n<a \le 2n}_{1\le \tau\le t}}\,.
\label{eq:basisrewriting}
\ee
We call these two copies of $\mathcal H_{nt}$ the ``positive-time'' and ``negative-time'' spaces respectively, as the components of $\mathbb T_{\theta,\phi}[h]$ and $\mathbb R_{\theta,\phi}[h]$ acting on those spaces come from terms in \eqref{eq:trrhoA} respectively propagating forward and backward in time. 

It is useful to note that $\mathbb T_{\theta,\phi}[h]$ and $\mathbb R_{\theta,\phi}[h]$ are the same up to a cyclic permutation of the copies of $\mathcal H_t$ composing the negative-time space (i.e. a cyclic permutation of the second row of Fig.~\ref{fig:tensorproductspace}), namely  
\be
\mathbb R_{\theta,\phi}[h] = \mathbb P\, \mathbb T_{\theta,\phi}[h]\, \mathbb P^{\dag}\,,
\ee
where we defined 
\be
\mathbb P =  \1\otimes\prod_{\nu=1}^{n}\prod_{\tau=1}^{t}  P_{(\nu,\tau),(\nu-1,\tau)}\,.
\label{eq:defPP}
\ee
Here $P_{(\nu,\tau),(\nu-1,\tau)}$ is an elementary transposition 
\be
P_{(\nu,\tau), (\nu',\tau')}= \frac{1}{2}\1+\frac{1}{2}\sum_{a\in\{x,y,z\}}\sigma^a_{\nu,\tau}\sigma^a_{\nu',\tau'}\,.
\label{eq:elementarytransposition}
\ee
With $\nu,\nu'\in\{1,\ldots,n\}$ and $\tau,\tau'\in\{1\ldots,t\}$. The matrix $\sigma^a_{\nu,\tau}$ acts as the Pauli matrix $\sigma^{a}$, $a\in\{x,y,z\}$, at the site $\tau=1,\ldots, t$ of the $\nu$-th copy of $\mathcal H_t$ in $\mathcal H_{nt}$, i.e. 
\be
\!\!\![\sigma^a_{\nu,\tau},\sigma^b_{\nu',\tau'}]=2 i \delta_{\nu,\nu'}\delta_{\tau,\tau'}\varepsilon^{abc}\sigma^c_{\nu,\tau}\,,\qquad \sigma^a_{0,\tau}\equiv\sigma^a_{n,\tau}. 
\ee
Note that the property $P_{(\nu,\tau),(\mu,\sigma)} = P_{(\nu,\tau),(\mu,\sigma)}^{-1}$ implies $\mathbb P^\dagger = \mathbb P^{-1}$.

Writing \eqref{eq:matrixelementsT} in matrix form we have that the transfer matrix is a simple tensor product of single-copy transfer matrices
\be
\mathbb T^{\phantom{(\nu)}}_{\theta,\phi}[h] = \prod_{\nu=1}^n\mathbb T^{(\nu)}_{\theta,\phi}[h] \,.
\label{eq:dualtranfermatrixentropy}
\ee
The matrix $\mathbb T^{(\nu)}_{\theta,\phi}[h]$ acts non-trivially only on the $\nu$-th copy of $\mathcal H_t$ in both the positive-time and negative-time spaces ($\nu$-th column of Fig.~\ref{fig:tensorproductspace}), and it is explicitly written as 
\begin{align}
\mathbb T^{(\nu)}_{\theta,\phi}[h] = \mathbb B^{z}_{\nu, 1}[\theta]\cdot \mathbb G^{z}_{\nu, t}\cdot \mathbb U_{\phi}^{(\nu)}[h]\,, 
\label{eq:productformT}
\end{align}
where we introduced the Hermitian matrix $\mathbb B^{a}_{\nu, \tau}[\theta]$, the projector $\mathbb G^{a}_{\nu, \tau}$, and the unitary matrix $\mathbb U_{\phi}^{(\nu)}[h]$ defined as follows 
\begin{align}
\mathbb B^{a}_{\nu, \tau}[\theta]&\equiv  2 \left[ \cos(\theta/2)  P_{\nu,\tau}^{a,+} +\sin(\theta/2)  P_{\nu,\tau}^{a,-}\right]^{\otimes 2}
,\\
\mathbb G^{a}_{\nu, \tau}&\equiv \frac{1}{2}(\1 + \sigma^a_{\nu, \tau}\otimes\sigma^a_{\nu, \tau}),\\
\mathbb U_{\phi}^{(\nu)}[h]&\equiv  
{U}_{\nu,\phi} e^{-i h  M^z_{\nu}} e^{i\frac{\pi}{4}  M^x_{\nu}}  \otimes {U}_{\nu,\phi}^* e^{i h M^z_{\nu}} e^{-i\frac{\pi}{4} M^x_{\nu}}\,,\label{eq:defUU}
\end{align}
and finally 
\begin{align}
{U}_{\nu,\phi} &\equiv \exp\left[-\frac{i\pi}{4} \sum_{\tau=1}^{t-1} \sigma^{z}_{\nu, \tau}  \sigma^z_{\nu, \tau+1}-i\frac{\phi}{2}\sigma^{z}_{\nu,1}\right]\!,\label{eq:defU}\\
{M}^a_{\nu} &\equiv  \sum_{\tau=1}^{t} \sigma^{a}_{\nu, \tau}\,,\label{eq:defM}\\
P_{\nu,\tau}^{a,\pm} &\equiv \frac{1}{2}( \1 \pm  \sigma^a_{\nu,\tau})\,.
\end{align}
Note that since 
\be
\left[\mathbb T^{(\nu)}_{\theta,\phi}[h] ,\mathbb T^{(\mu)}_{\theta,\phi}[h] \right]=0\,,\qquad\qquad\mu,\nu\in\{1,\ldots,n\},
\ee
the order in the product \eqref{eq:dualtranfermatrixentropy} is irrelevant.

\begin{widetext}
Putting everything together we have 
\be
S^{(n)}_A(t)=\frac{1}{1-n}\log {\rm tr}\left[\left(\prod_{j=1}^N \mathbb T_{\theta_j,\phi_j}[h_j]\right)\mathbb P\,  \left (\prod_{j=N+1}^L \mathbb T_{\theta_{j},\phi_{j}}[h_{j}]\right)\mathbb P^{\dag}\right]\!.
\label{eq:entropyduality}
\ee
\end{widetext}
This equation accomplishes the duality mapping of the entanglement entropies: we wrote the entanglement entropies in terms of the trace of products of an appropriate transfer matrix in space. 

Before continuing with the evaluation of \eqref{eq:entropyduality} two comments are in order. First we note that the mapping described can be performed also when $J$ and $b$ in \eqref{eq:hamiltonians1} and \eqref{eq:hamiltonians2} do not fulfil the self-duality condition \eqref{eq:selfduals}. For generic $J$ and $b$ we obtain that the entropy is still given by \eqref{eq:entropyduality} but the transfer matrix $ \mathbb T_{\theta,\phi}[h]$ is modified in two ways. (i) the matrix $\mathbb U_{\phi}^{(\nu)}[h]$ is not unitary anymore. The Ising coupling in \eqref{eq:defU} replaced by $\tilde J$ (\emph{cf}. \eqref{eq:tildeJ}) and the transverse fields in \eqref{eq:defUU} (the coefficients of $i M^x_{\nu}$ in the positive and negative time copy) are respectively replaced by $\tilde b$ and $\tilde b^*$ (\emph{cf}. \eqref{eq:tildeb}). (ii) the projector $\mathbb G^{z}_{\nu, t}$ in \eqref{eq:entropyduality} is replaced by 
\be
\cos\left(b\left(\sigma^z_{\nu, t}\otimes\1-\1\otimes\sigma^z_{\nu, t}\right)\right)\,.
\ee  
As we will see in the next section these changes are enough to hinder the analytical evaluation of \eqref{eq:entropyduality}, however, the duality approach can still be useful for perturbative calculations or numerical approaches (see Sec.~\ref{sec:genericcase}).

We also observe that when the initial state is in the class $\cal L$ (\emph{cf}.~\eqref{eq:classesL}), namely when 
\be
\theta_{j}=\bar\theta_j\in\{0,\pi\},\qquad j \in\{1,2,\ldots,L\},
\ee
the expression \eqref{eq:entropyduality} can be further simplified by effectively reducing the dimension of the space where the trace acts. This is explicitly shown in Appendix~\ref{app:longstates}. The final result is again of the form \eqref{eq:entropyduality} with the replacement  
\be
\begin{split}
{\mathbb T}_{\bar \theta_j,\phi_j}[h_j] &\longmapsto \bar{\mathbb T}_{\frac{\pi}{2},\bar \theta_j-\frac{\pi}{2}}[h_j]\,,\\
{\mathbb P} &\longmapsto \bar{\mathbb P}. 
\end{split}
\label{eq:TtoTbar}
\ee
Here the barred operators have exactly the same form as the non-barred ones (respectively \eqref{eq:dualtranfermatrixentropy} and \eqref{eq:defPP}) but act on $\mathcal H_{t-1}^{\otimes 2n}$ instead of $\mathcal H_{t}^{\otimes 2n}$. Note that this is nothing but a restatement of Property~\eqref{eq:LtoT}.

%%%%%%%%%%%%%%%%%%%%%%%%%%%%
\section{Separating states}%
%%%%%%%%%%%%%%%%%%%%%%%%%%%%
\label{sec:separating}

Our goal is to use Equation~\eqref{eq:entropyduality} to determine $S^{(n)}_A(t)$ in the thermodynamic limit. To do that, however, we need some information on the Jordan normal form of the matrix $\mathbb T_{ \theta,\phi}[h]$. Indeed, since the matrix is not normal, it is not guaranteed to be (and it is generically not) diagonalisable.

As proven in Appendix~\ref{app:proofsp1p2}, the form \eqref{eq:dualtranfermatrixentropy}--\eqref{eq:productformT} of the transfer matrix has some simple but useful consequences on its Jordan normal form. Specifically we have   
\begin{property}
\label{prop:transverse}
The following facts hold
\begin{itemize}
\item[(i)] $|\lambda_j|\leq \lambda_{\rm max}\equiv (1+|\cos\theta|)^n$, $\forall\lambda_j \in {\rm Spec}\left[\mathbb T_{ \theta,\phi}[h]\right].$
\item[(ii)] If an eigenvalue $\lambda$ of $\mathbb T_{\theta,\phi}[h]$ fulfils $|\lambda|=\lambda_{\rm max}$ then
\begin{itemize}
\item[a.] $\lambda$ has trivial Jordan blocks (its geometric and algebraic multiplicities coincide).
\item[b.] the associated left eigenvector $\bra{A}$ satisfies 
\begin{align}
\bra{A}\prod_{\nu=1}^n  \mathbb B^{z}_{\nu, 1}[\theta]&= \lambda_{\rm max} \bra{A},
\label{eq:eigproj1}\\
\bra{A}\prod_{\nu=1}^n \mathbb G^{z}_{\nu, t} &=\bra{A}, \label{eq:eigproj}\\
\bra{A}\prod_{\nu=1}^n \mathbb U_{\phi}^{(\nu)}[h]&= e^{i \alpha} \bra{A},\quad \alpha\in\mathbb R\,,
\label{eq:eigTT}
\end{align}
\end{itemize}
\end{itemize}
\end{property}
\noindent where ${\rm Spec}\left[A\right]$ denotes the spectrum of the matrix $A$. 

Property~\ref{prop:transverse} introduces the crucial simplification of this work. If the maximal eigenvalues of $\mathbb T_{\theta,\phi}[h]$ saturate the bounds at point $(i)$ the problem of finding the maximal eigenvalues of the transfer matrix is \emph{separated} into three much easier ones, consisting of finding eigenvalues and eigenvectors of simple hermitian and unitary matrices. 

The bound at point $(i)$, however, cannot be always saturated. To see this, let us consider some constrains on the structure of the matrix $\mathbb T_{\theta,\phi}[h]$ coming from the identity~\eqref{eq:traceintermediate}. These are most easily found by considering the translational invariant case 
\be 
h_j=h,\qquad\theta_j=\theta,\qquad \phi_j=\phi,\qquad \forall j\,.
\ee
Setting $N=0$ in \eqref{eq:traceintermediate} we have 
\be
{\rm tr}\!\!\left[ \left(\mathbb T_{\theta,\phi}[h]\right)^L\right]={\rm tr}\left[(\rho(t))^n\right]=1,\qquad \forall\, L,n\,,
\ee
where in the second step we used that the state \eqref{eq:statet} is pure. This relation implies that the eigenvalues of $\mathbb T_{ \theta,\phi}[h]$ are all 0 but one, which is equal to 1. Moreover, the Jordan block corresponding to the eigenvalue 1 is one-dimensional, while the eigenvalue 0 might have (and does have!) a highly nontrivial Jordan structure. More formally:
\be
\label{eq:physicalrequest}
\begin{split}
({\rm C}1)&\;\; {\rm Spec}\left[\mathbb T_{ \theta,\phi}[h]\right]=\{0,1\}\,.\\
({\rm C}2)&\;\; {\rm The\;geometric\;multiplicity}\\
&{\rm \;of\; the\;eigenvalue\;1\;is\;1}.
\end{split}
\ee
From the conditions \eqref{eq:physicalrequest} it follows that the bound at point $(i)$ of Property~\ref{prop:transverse} can be saturated only when $\lambda_{\rm max}=1$. Note that the cases for which $\lambda_{\rm max}=1$ include ${\theta={\pi}/2}$, but also $\theta=0,\pi$. Indeed in the latter case the matrix $\mathbb T_{\theta,\phi}[h]$ can be replaced by $\bar{\mathbb T}_{{\pi}/2,\pi/2-\theta}[h]$ (see \eqref{eq:TtoTbar} and Appendix~\ref{app:longstates}). In other words, the requirement $\lambda_{\rm max}=1$ selects the two classes of states ${\cal T}$ and ${\cal L}$ introduced in Sec.~\ref{sec:results}. This clarifies the meaning of their name. We called them ``separating" states because if the initial state is one of them the problem of finding the maximal eigenvalues of the transfer matrix (and the corresponding eigenvectors) can be separated. In the upcoming section we explicitly solve the separated problem \eqref{eq:eigproj1}--\eqref{eq:eigTT} for $\lambda_{\max}= 1$, and, incidentally, we also verify that it has no solution for $\lambda_{\max}\neq 1$. 

Finally, we note that away from the self dual points \eqref{eq:selfduals} the conditions \eqref{eq:physicalrequest} still hold and a property similar to Property~\ref{prop:transverse} is still valid. In that case, however, the bound can never be saturated and no separation can be performed. This makes the problem analytically intractable, at least in an exact fashion.

\section{Entanglement spreading from separating states}
\label{sec:specialcase}

Here we explicitly solve the entanglement evolution from separating states. In particular in Sec.~\ref{sec:eigenvector} we solve the separated problem \eqref{eq:eigproj1}--\eqref{eq:eigTT}  for $\lambda_{\max}=1$ and in Sec.~\ref{sec:entdynamics} we evaluate \eqref{eq:entropyduality}. To be concrete we focus on initial states in the class $\mathcal T$, the result for states in the class $\mathcal L$ is obtained using \eqref{eq:LtoT}.  

\subsection{Maximal eigenvalues of the transfer matrix}
\label{sec:eigenvector}

Our strategy is to determine the maximal eigenvalues of $\mathbb T_{\pi/2,\phi}[h]$ and the associated eigenvectors, by searching for all the vectors fulfilling \eqref{eq:eigproj1}--\eqref{eq:eigTT} with $\lambda_{\max}=1$. To simplify our analysis we make two observations. First, we note that 
\be
\mathbb B^{z}_{\nu, 1}[\tfrac{\pi}{2}]=\1\otimes\1\,,
\ee
so that \eqref{eq:eigproj1} becomes trivial. Second, we note that all $\mathbb G^{z}_{\nu,t}$ and $\mathbb U_{\phi}^{(\nu)}$ commute for different $\nu$s so we can look for simultaneous eigenvectors. The problem is then reduced to finding all vectors $\bra{A}$ fulfilling  
\begin{align}
 \bra{A} \mathbb G^{z}_{\nu,t}=&\bra{A}\,,\label{eq:condAs1}\\
  \bra{A}\mathbb U_{\phi}^{(\nu)}=& \bra{A} e^{i \alpha_\nu}\,,\quad \alpha_\nu\in\mathbb R\,,\quad\forall\nu\in\{1,\ldots,n\}\,.\label{eq:condAs2}
\end{align} 
To solve these equations it is convenient to introduce the following one-to-one vector-to-operator mapping (\emph{cf}.  Ref.~\cite{letter}) $\bra{A} \leftrightarrow A$:
\be
\bra{A}={\sum_{k,m}} \bra{m}A\ket{k} \bra{k}\otimes \bra{m}^*\,,
\label{eq:statetoop}
\ee 
where $\{\bra{k}\}$ is a basis of $\mathcal H_{n t}$ and $(\cdot)^*$ denotes complex conjugation in the computational basis $\mathcal B_{nt}$, such that 
\be
\bra{k}^*O^*\ket{m}^*=\braket{k|O|m}^*\!\!, 
\label{eq:complexconjugate}
\ee
for any operator $O$. Using the mapping \eqref{eq:statetoop}, Eqs.~\eqref{eq:condAs1}--\eqref{eq:condAs2} are directly rewritten in operatorial form as follows  
\begin{align}
\sigma^z_{\nu,t}\, A&= A\, \sigma^z_{\nu,t}\,,\label{eq:condAo1}\\
{U}_{\nu,\phi}\,  e^{-i h  M_\nu^{z}}\, e^{i\frac{\pi}{4}  M_\nu^x}  A &= e^{i \alpha_\nu}  A\, {U}_{\nu,\phi}\, e^{-i h  M_\nu^z}\,e^{i\frac{\pi}{4}  M_\nu^x},
\label{eq:condAo2}
\end{align}
for some $\alpha_{\nu}\in\mathbb R$ and all $\nu\in\{1,\ldots,n\}$. In this formulation our goal is to find all independent linear operators $A$ over ${\cal H}_{nt}$ solving the commutation relations \eqref{eq:condAo1}--\eqref{eq:condAo2}. As shown in Appendix~\ref{app:proofsp3p4}, these commutation relations are equivalent to     
\be
 A \sigma^a_{\nu,\tau} = \sigma^a_{\nu,\tau} A\,, 
 \label{eq:basiccommutation}
\ee
for all $a\in\{x,y,z\}$, $\tau\in\{1,\ldots, t\}$, $\nu\in\{1,\ldots,n\}$. Namely, they are equivalent to requiring that $A$ commutes with the entire algebra of observables in $\mathcal H_{ n t}$. Since the latter is irreducible, Shur's Lemma implies that the unique (up to multiplicative factors) solution to \eqref{eq:basiccommutation} is given by 
\be
A=\1\,,\quad \text{and}\quad \alpha_\nu=0\,.
\ee
We then find that the eigenvalue of $\mathbb T_{\pi/2, \phi}[h]$ with maximal magnitude is $1$ and corresponds to the unique left eigenvector
\be
\bra{\1}=\frac{1}{2^{n t /2}}\sum_{\{s_{\nu,\tau}\}} \bra{\{s_{\nu,\tau}\}}\otimes \bra{\{s_{\nu,\tau}\}}\,,
\ee
where we used the computational basis, omitted complex conjugation as the basis is real, and we included the normalisation factor ${\sqrt{{\rm tr}[\1]}=2^{n t/2}}$. Note that the unique right eigenvector of $\mathbb T_{\pi/2, \phi}[h]$ associated to ${\lambda=1}$ is given by $\ket{\1}=(\bra{\1})^\dag$, as it can be directly verified.

We also observe that since we just proved that \eqref{eq:eigproj} and \eqref{eq:eigTT} have $\bra{A}=\bra{\1}$ as only solution, and, moreover 
\be
\bra{\1}\prod_{\nu=1}^n  \mathbb B^{z}_{\nu, 1}[\theta] \neq \lambda_{\rm max} \bra{\1}, \qquad \theta\neq\pi/2\,,
\ee
the separated problem \eqref{eq:eigproj1}--\eqref{eq:eigTT} has no solution for ${\theta\neq\pi/2}$. 

\subsection{Entanglement dynamics}
\label{sec:entdynamics}

Our next step is to use the eigenvectors determined above to compute the entanglement dynamics. First we note that the eigenvector $\ket{\1}$ is independent of $\phi$ and $h$. Moreover, $\ket{\1}$ is orthogonal to all left generalised eigenvectors corresponding to the eigenvalues $0$ of $\mathbb T_{\pi/2, \phi}[h]$ for all $\phi$ and $h$. These two facts imply 
\be
\!\!\!\lim_{L\rightarrow \infty} S^{(n)}_A(t)=\frac{1}{1-n}\log\!\!\left[\!\braket{\Psi|\!\!\left(\prod_{j=1}^N \mathbb T_{{\pi}/{2},\phi_j}[h_j]\right)\!\!|\Psi}\!\right]\!\!,
\label{eq:TLentropysepinhomo}
\ee
where we introduced  
\be
\ket{\Psi}\equiv\mathbb P^{\dag}\ket{\1},\qquad\bra{\Psi}\equiv(\ket{\Psi})^{\dag}=\bra{\1}\mathbb P\,.
\label{eq:psi}
\ee
The relation \eqref{eq:TLentropysepinhomo} can be used to find the slope of the linear growth of the entanglement entropy. Indeed, taking $N$ to infinity we have 
\be
\lim_{N\to\infty}\lim_{L\rightarrow \infty} S^{(n)}_A(t)=\frac{2}{1-n}\log|\!\braket{\Psi|\1}\!|=2t\log2\,.
\ee
The simple structure of $\mathbb T_{{\pi}/{2},\phi}[h]$, however, allows us to progress further and evaluate \eqref{eq:TLentropysepinhomo} exactly for each $N$. 

\begin{widetext}
This can be done by making use of the following remarkable identity 
\be
\braket{\Psi|\left(\prod_{j=1}^N \mathbb T_{\tfrac{\pi}{4},\phi_j}[h_j]\right)|\Psi}=\braket{\Psi|\prod_{\nu=1}^{n}\left[\prod_{\tau=0}^{\lfloor \frac{N}{2}\rfloor-1}\!\!\!\!\!\left[\mathbb G^{z}_{\nu, t-\tau}\mathbb G^{x}_{\nu, t-\tau}\right]\left[\mathbb G^{z}_{\nu, t-\lfloor \frac{N}{2}\rfloor}\right]^{{\rm mod}(N,2)}\right]|\Psi}\,,\qquad\forall \phi_j,h_j\,,
\label{eq:identity}
\ee
where $\lfloor\cdot\rfloor$ denotes the floor function. Here we adopted the convention 
\be
\mathbb G^{a}_{\nu, \tau}=\1\,,\qquad\qquad \tau\leq0\,,
\ee
and, to lighten the notation, from now on we assume that a product $\prod \cdots$ only picks a single factor on its right unless several terms are grouped within a square bracket $[\cdots]$. 
\end{widetext} 

The identity \eqref{eq:identity} is proven in Appendix~\ref{app:proofoffund} using the explicit form of $\mathbb T_{{\pi}/{2},\phi}[h]$ and the following useful properties of the state \eqref{eq:psi}
\begin{align}
&\prod_{\nu=1}^{n} O_\nu\otimes O^{*}_\nu \ket{\Psi}=\ket{\Psi}\,,
\label{eq:stateproperty}\\
&\bra{\Psi} \prod_{\nu=1}^{n} O_\nu\otimes O^{*}_\nu =\bra{\Psi}\,,
\label{eq:statepropertyleft}
\end{align}
where $O_\nu$ acts non trivially, as the unitary operator $O$, only on the $\nu$-th copy of $\mathcal H_t$ in $\mathcal H_{nt}$, i.e. 
\be
O_\nu = \one_{{\cal H}_t}^{\otimes(\nu-1)}\otimes O\otimes \1_{{{\cal H}_t}}^{\otimes (n-\nu)}.
\ee
These properties are proven in Appendix~\ref{app:proofofstateproperty}. 

A striking consequence of \eqref{eq:identity} is that the entanglement entropies evolving from separating states are completely independent of the configuration of longitudinal magnetic fields $\{h_j\}$ and of the initial-state angles $\{\phi_i\}$. For instance, this means that the same result is obtained in the integrable and in the non-integrable case, with or without disorder. 

The evaluation of the r.h.s. of \eqref{eq:identity} is now straightforward. First, we note that in the computational basis \eqref{eq:basisrewriting} of $\mathcal H_{t}^{\otimes 2n}$ we have 
\begin{widetext}
\begin{align}
&\bra{\{ s'_{\nu,\tau}\}}\otimes\bra{\{r'_{\nu,\tau}\}}\prod_{\nu=1}^{n}\left[\prod_{\tau=0}^{\lfloor\! \frac{N}{2}\!\rfloor-1}\!\!\!\!\!\left[\mathbb G^{z}_{\nu, t-\tau}\mathbb G^{x}_{\nu, t-\tau}\right][\mathbb G^{z}_{\nu, t-\lfloor\! \frac{N}{2}\!\rfloor}]^{{\rm mod}(N,2)}\right]\ket{\{s_{\nu,\tau}\}}\otimes\ket{\{r_{\nu,\tau}\}}\notag\\
 &=
 \begin{cases}
\displaystyle \frac{1}{2^{n\lfloor\! \frac{N}{2}\!\rfloor}} \prod_{\nu=1}^{n}
\left[\prod_{\tau=0}^{t-\lfloor\! \frac{N}{2}\!\rfloor}\!\!\left[\delta_{r'_{\nu,\tau} r^{\phantom{\prime}}_{\nu,\tau}}\delta_{s'_{\nu,\tau} s^{\phantom{\prime}}_{\nu,\tau}}\right] \!\!
\Bigl[\delta_{s_{\nu,t-\lfloor\! \frac{N}{2}\!\rfloor},r_{\nu,t-\lfloor\! \frac{N}{2}\!\rfloor}}\Bigr]^{{\rm mod}(N,2)} \!\!\!\!\!\!\!\!\!\!\prod_{\tau=t-\lfloor\! \frac{N}{2}\!\rfloor+1}^{t}\!\!\left[\delta_{s^{\phantom{\prime}}_{\nu,\tau} r^{\phantom{\prime}}_{\nu,\tau}}\delta_{s'_{\nu, \tau},r'_{\nu,\tau}}\right]
 \right]  & \lfloor\! \frac{N}{2}\!\rfloor< t\\
\\
\displaystyle \frac{1}{2^{n t}} \prod_{\nu=1}^{n} \prod_{\tau=1}^{t}\left[\delta_{s^{\phantom{\prime}}_{\nu,\tau},r^{\phantom{\prime}}_{\nu,\tau}}\delta_{s'_{\nu, \tau},r'_{\nu, \tau}}\right]  & \lfloor\! \frac{N}{2}\!\rfloor\geq t
 \end{cases}\,,
 \label{eq:EVcomputationalbasis}
\end{align}
where the matrix elements of $\mathbb G^{x}_{\nu, \tau}$ and $\mathbb G^{z}_{\nu, \tau}$ are computed by repeated use of
\begin{align}
    & \bra{s'}\otimes\bra{r'}\,\one\, \ket{s}\otimes\ket{r} = \delta_{s,s'}\delta_{r,r'}\,,\notag\\
    & \bra{s'}\otimes\bra{r'}\frac{1}{2}\left(\one+\sigma^z\otimes\sigma^z\right)\ket{s}\otimes\ket{r} = \delta_{s,s'}\delta_{r,r'}\delta_{s,r}=  \delta_{s,s'}\delta_{s,r}\delta_{s',r'}\,,\notag\\
   & \bra{s'}\otimes\bra{r'}\frac{1}{2}\left(\one+\sigma^z\otimes\sigma^z\right)\frac{1}{2}\left(\one+\sigma^x\otimes\sigma^x\right)\ket{s}\otimes\ket{r} = \frac{1}{2}\delta_{s,r}\delta_{s',r'}\,,\qquad\qquad\qquad s,r,s',r'\in\{\pm1\}\,.
\end{align}

Then, we plug \eqref{eq:EVcomputationalbasis} into the r.h.s. of Eq.~\eqref{eq:identity}. For $t> \lfloor\! N/2\!\rfloor$ we find 
\begin{align}
&\braket{\Psi|\prod_{\nu=1}^{n}\left[ \prod_{\tau=0}^{\lfloor\! \frac{N}{2}\!\rfloor-1}\!\!\!\left[\mathbb G^{z}_{\nu, t-\tau}\mathbb G^{x}_{\nu, t-\tau}\right][\mathbb G^{z}_{\nu, t-\lfloor\! \frac{N}{2}\!\rfloor}]^{{\rm mod}(N,2)}\right]|\Psi}=\notag\\
&= \frac{1}{2^{n \lfloor\! \frac{N}{2}\!\rfloor+n t}} \sum_{\{s^{\phantom{\prime}}_{\nu,\tau}\}} \sum_{\{s^{{\prime}}_{\nu,\tau}\}}
\prod_{\nu=1}^{n}
\left[\prod_{\tau=1}^{t-\lfloor\! \frac{N}{2}\!\rfloor}\!\!\left[\delta_{s'_{\nu+1,\tau} s^{\phantom{\prime}}_{\nu+1,\tau}}\delta_{s'_{\nu,\tau} s^{\phantom{\prime}}_{\nu,\tau}}\right] 
\left[\delta_{s_{\nu,t-\lfloor\! \frac{N}{2}\!\rfloor},s_{\nu+1,t-\lfloor\! \frac{N}{2}\!\rfloor}}\right]^{{\rm mod}(N,2)} \!\!\!\!\!\!\prod_{\tau=t-\lfloor\! \frac{N}{2}\!\rfloor+1}^{t}\!\!\left[\delta_{s^{\phantom{\prime}}_{\nu,\tau} s^{\phantom{\prime}}_{\nu+1,\tau}}\delta_{s'_{\nu, \tau},s'_{\nu+1,\tau}}\right]
 \right] \notag\\
&= \frac{1}{2^{n\lfloor\! \frac{N}{2}\!\rfloor+n t}} \left[\sum_{\{s_{\nu,\tau}\}_{\tau < t-\lfloor\! \frac{N}{2}\!\rfloor}} 1 \right] \left[\sum_{\{s_{\nu,t-\lfloor\! \frac{N}{2}\!\rfloor}\}} \left[\delta_{s_{\nu,t-\lfloor\! \frac{N}{2}\!\rfloor},s_{\nu+1,t-\lfloor\! \frac{N}{2}\!\rfloor}}\right]^{{\rm mod}(N,2)} \right] \left[\sum_{\{s_{1,\tau}\}_{\tau> t-\lfloor\! \frac{N}{2}\!\rfloor}} 1 \right]^2  \notag\\
&= \frac{1}{2^{n \lfloor\! \frac{N}{2}\!\rfloor +n t}}   2^{n(t-\lfloor\! \frac{N}{2}\!\rfloor-1)} 2^{n-(n-1){\rm mod}(N,2)} 2^{2 \lfloor\! \frac{N}{2}\!\rfloor} =2^{N(1-n)}\,.
 \end{align}
\end{widetext}
Proceeding analogously for $t\leq \lfloor N/2\rfloor$ we have  
\be
\!\!\!\!\!\braket{\Psi|\!\!\!\!\prod_{\tau=0}^{\lfloor\! \frac{N}{2}\!\rfloor-1}\!\!\!\!\!\left[\mathbb G^{z}_{\nu, t-\tau}\mathbb G^{x}_{\nu, t-\tau}\right]\!\!\bigl[\mathbb G^{z}_{\nu, t-\lfloor\! \frac{N}{2}\!\rfloor}\bigr]^{\!{\rm mod}(N,2)}\!|\Psi}\!\! = \!{2^{2t(1-n)}}\!.
\ee
Therefore, we finally obtain that for initial states in the class $\mathcal T$, all entanglement entropies $S^{(n)}_A(t)$ with ${n=2,3,\ldots}$ are exactly given by Eq.~\eqref{eq:finalresult}. This, however, implies that 
\be
{\rm Spec}\left[\rho_A(t)\right] =  \left\{2^{-\min(2t,N)}, 0 \right\}\,,
\ee
where $2^{-\min(2t,N)}$ has multiplicity $2^{\min(2t,N)}$, while $0$ has multiplicity $2^{N-\min(2t,N)}$. As a consequence, the result \eqref{eq:finalresult} holds for all $S^{(\alpha)}_A(t)$ with any {\em real positive} $\alpha$.

\begin{figure*}[t]
\includegraphics[width=0.90\textwidth]{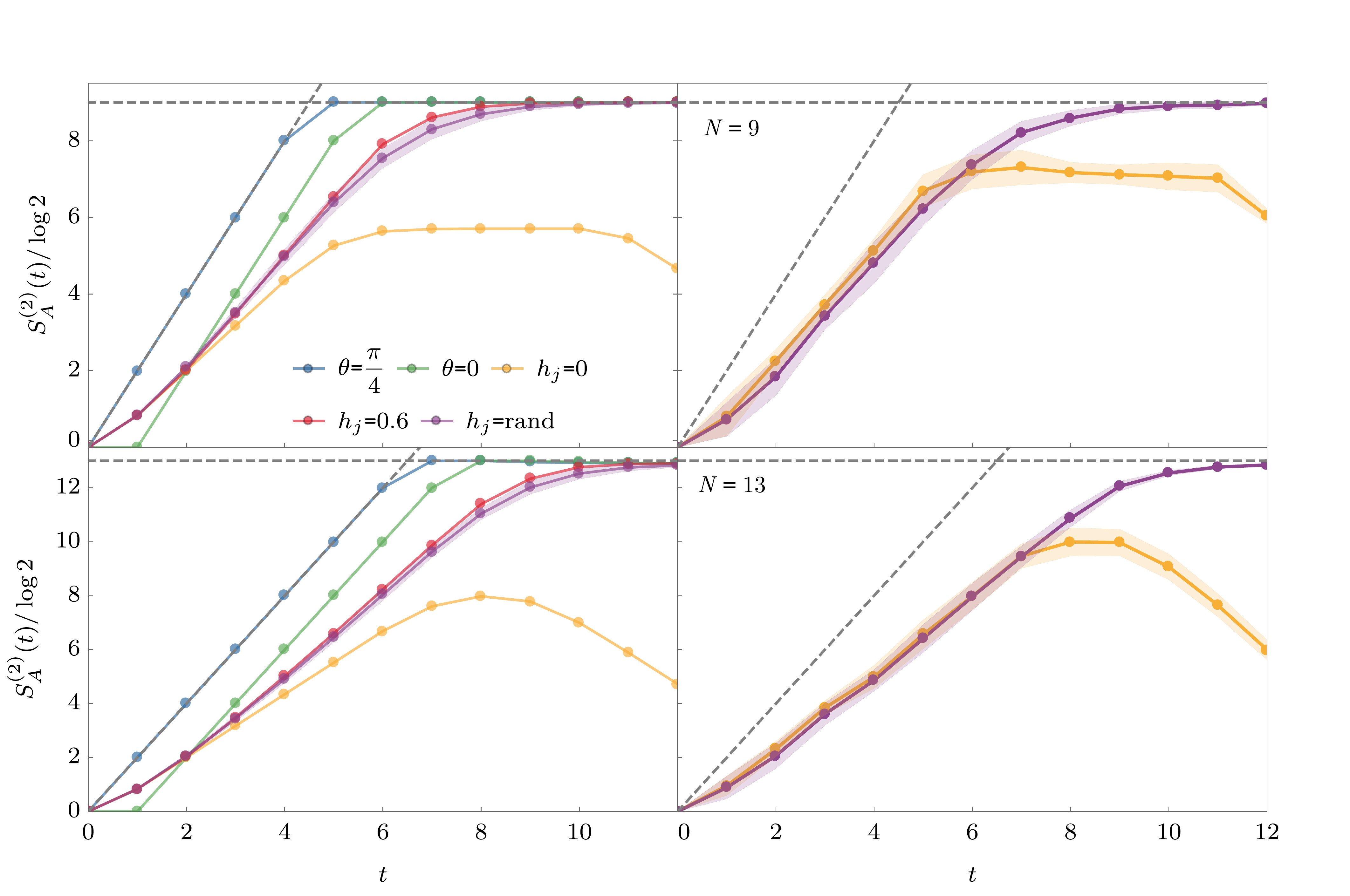}
\caption{%
The second R\'enyi entropy for a kicked Ising system of $L=30$ spins evolving from ``tilted'' initial states \eqref{eq:state0}. Top and bottom two panels have respectively $N=9$ and $N=13$. The two panels on the left report results for translational invariant initial states. The blue and green curves correspond respectively to transverse and longitudinal separating states (\emph{cf.} \eqref{eq:classesT} and \eqref{eq:classesL}). Other curves correspond to the initial state $\theta_j=\phi_j=1$ and different magnetic fields as indicated in the legend. The two panels on the right correspond to the maximally disordered cases, where the spins at each site point in a random direction and the magnetic fields $h_j$ are either random (purple) or zero (yellow). 
In the cases with random parameters we show the average values for a sample of 8 realisations using a continuous line and indicate a standard deviation of one realisation by a shaded area.}
\label{fig:DiffStates}
\end{figure*}

%%%%%%%%%%%%%%%%%%%%%%%%%%%%
\section{Entanglement spreading from generic states}%
%%%%%%%%%%%%%%%%%%%%%%%%%%%%
\label{sec:genericcase}

The exact results derived in the previous sections have three remarkable features. (i) The entropies do not depend at all on the longitudinal magnetic fields. In particular, they are not affected by whether or not the system is integrable; (ii) The entropies grow at the maximal speed allowed by the range of the Hamiltonian and the dimension of the local Hilbert space (they saturate the minimal cut bound \eqref{eq:mincut}); (iii) At each fixed time $t$ all entanglement entropies coincide, signalling a flat entanglement spectrum, \emph{i.e.}, that all non-zero eigenvalues of the density matrix reduced to the block $A$ are equal.

It is interesting to wonder whether these are general features of the entanglement spreading in the self-dual kicked Ising chain or, instead, they are special properties of separating initial states. In other words, it is interesting to ask whether the entanglement dynamics from separating states is an exceptional case or, even though special, it can be used to model the generic behaviour. To this aim, in this section we consider the entanglement spreading from generic product states \eqref{eq:state0} which are not separating. In this case, as pointed out above, we are unable to address the problem in a fully analytical fashion and we resort to a numerical analysis. 

From the physical point of view it is easy to see that the most convenient time regimes to examine possible modifications of the features (i)--(iii) are very different. Indeed, for ${\boldsymbol h \neq \boldsymbol 0}$ the system is ergodic, and any finite subsystem is expected to relax to the infinite temperature state irrespectively of the initial conditions. This means all entropies are expected to saturate to the universal value $N\log2$. On the other hand, for ${\boldsymbol h = \boldsymbol 0}$ the system is integrable and finite subsystems relax to generalised Gibbs ensembles~\cite{EF:GGE, VR:GGE}. We then expect the stationary values of the entropies to retain some memory of the initial configuration. To highlight the difference between integrable and non-integrable systems it is then convenient to focus on the ``saturation regime'' ${t\sim N}$, where the entropies become stationary. The generic relaxation to the infinite temperature state, however, also means that to see some dependence of the entanglement spectrum on $\boldsymbol h$, or on the initial state, one has to stay away from the saturation regime and focus on the ``growth regime'', ${t \ll N}$. The latter is obviously also the regime of interest to study variations in the speed of entanglement growth.

The saturation regime can be easily accessed by a ``direct" numerical approach. Namely, we consider a finite volume $L$ and determine the time evolving state by means of the efficient time-propagation algorithm described in the supplemental material of Ref.~\cite{letter}. The entanglement entropies are found by computing and diagonalising the reduced density matrices $\rho_A(t)$ (\emph{cf}. Eq.~\eqref{eq:reduceddensitymatrix}) and using Eq.~\eqref{eq:Renyi}. Note that a similar numerical analysis, in the case of the von Neumann entropy, has been performed in Ref.~\cite{PL:kickedIsing}.

\begin{figure*}[t]
\includegraphics[width=0.7\textwidth]{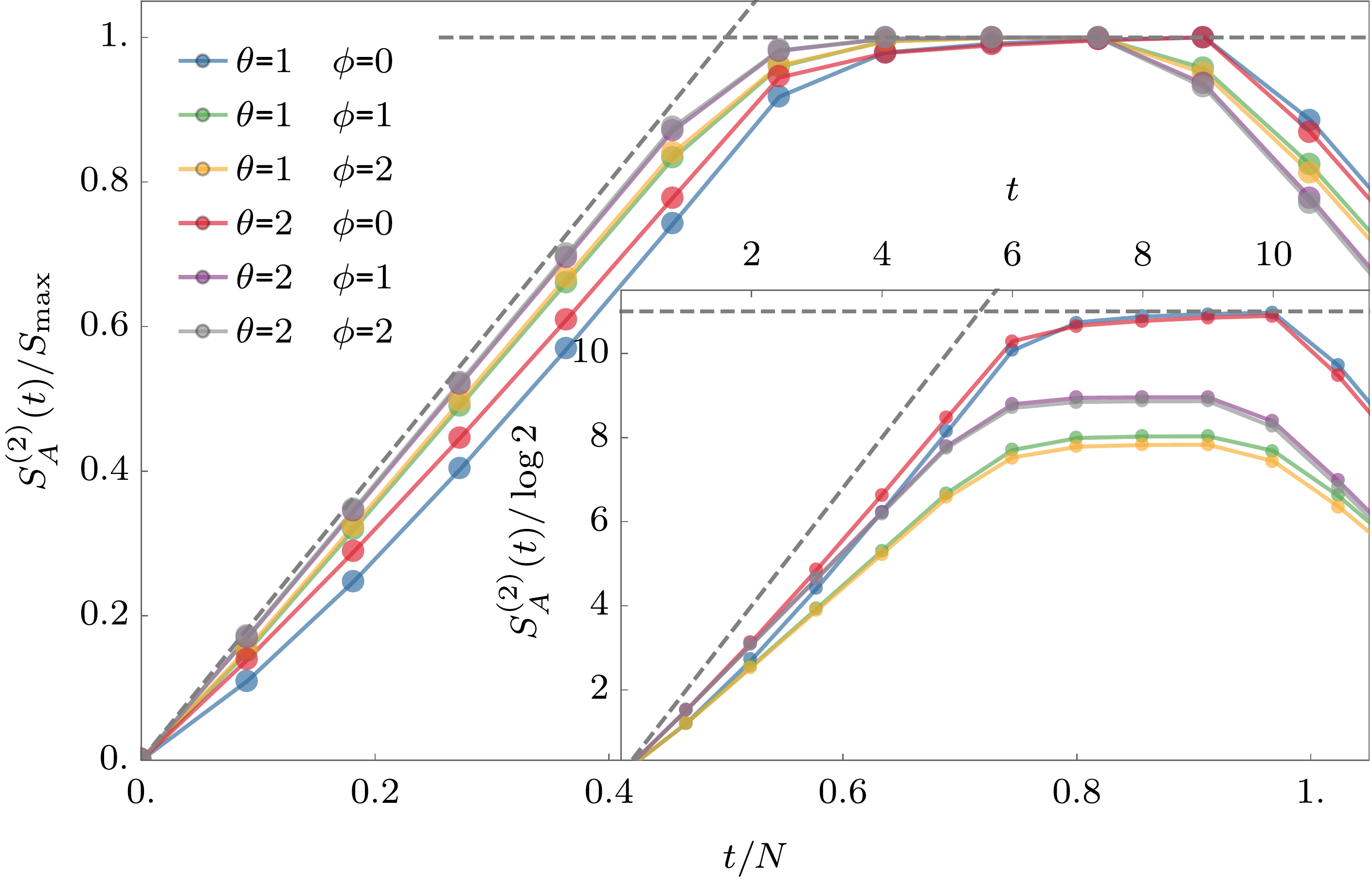}
\caption{Time evolution of the second R\'enyi entropy for a subsystem of $N=11$ spins in a kicked Ising system of $L=30$ at the integrable point $\boldsymbol h=0$ for different translationally invariant initial states. In the main shows the rescaled curves, which are close to  \eqref{eq:scalingfree} (black dashed line), given by the quasiparticle picture. In the inset we show the non-rescaled version, where it is apparent that the saturation value depends on the initial state. Note a recurrence after the time $t=10$, consistent with the quasiparticle picture.}
\label{fig:plotIntegrable}
\end{figure*}

Some representative examples of our results are reported in Fig.~\ref{fig:DiffStates}. First of all we see that the qualitative behaviour of the entanglement entropies is the same as for separating states, both in the 
homogeneous (translationally-invariant) and in the inhomogeneous case. The entropy grows in an approximately linear fashion until it saturates to a value proportional to the subsystem size. There is, however, a clear qualitative difference emerging between the integrable case and the generic one: in the generic case the entropies always saturate to $N \log2$ (minus the expected correction due to a finite $N/L$~\cite{Page, Lev, KH:NonIntEnt}), while this does not happen at the integrable point. In particular, in the inset of Fig.~\ref{fig:plotIntegrable} we report the evolution of $S_A^{(2)}(t)$ for several homogeneous non-separating initial states evolving under the integrable kicked Ising Hamiltonian. We see that, in contrast to the generic case, the saturation values depend on the initial state.  

Interestingly, we see that the evolution of the entropies shows very different finite-size effects in the integrable and non-integrable cases. In the former the entropies start to decrease at times larger than $(L-N)/2$, while in the latter they remain constant once they reached the saturation values. These behaviours  respectively agree with the predictions of the quasiparticle and the minimal-membrane picture. Indeed, for $L>2N$ the surface of the membrane is not affected by the system being finite. On the contrary, the quasiparticle picture predicts oscillations of the entropies due to quasiparticles traversing the entire system and going back to their initial positions. In particular, if the initial state is homogeneous, using that in our case the quasiparticles have all unit speed (and taking, for convenience, $L$ even), we find that the quasiparticle-picture prediction is $L/2$-periodic and, for $t\in\{0,1,\ldots,L/2\}$, it reads as 
\be
S_A^{(\alpha)}(t) = \min\left(2t,L-2t,N\right) S^{(\alpha)}_{\theta,\phi},
\label{eq:scalingfree}
\ee 
where $S^{(\alpha)}_{\theta,\phi}\leq \log 2$ is a ($N$- and $L$- independent) constant. This prediction holds in the asymptotic limit $t,N\to \infty$ with fixed $t/N$, but, as shown in the main panel of Fig.~\ref{fig:plotIntegrable}, it is in fair agreement with our numerical results already for $N=11$. Note that, even if the system is free, $S^{(\alpha)}_{\theta,\phi}$ cannot be generically computed analytically. Indeed, for generic values of $\boldsymbol \theta$ and $\boldsymbol \phi$ the states are not Gaussian in terms of the time evolving fermions and this makes the problem analytically untreatable. Moreover, since the dispersion is linear, the usual arguments about Gaussification do not apply \cite{Gaussification1,Gaussification2}. Interestingly, not even separating states are always Gaussian: transverse separating states are Gaussian only for $\phi_i=0,\pi$~\cite{notegaussian}.

\begin{figure*}[t]
\includegraphics[width=0.7\textwidth]{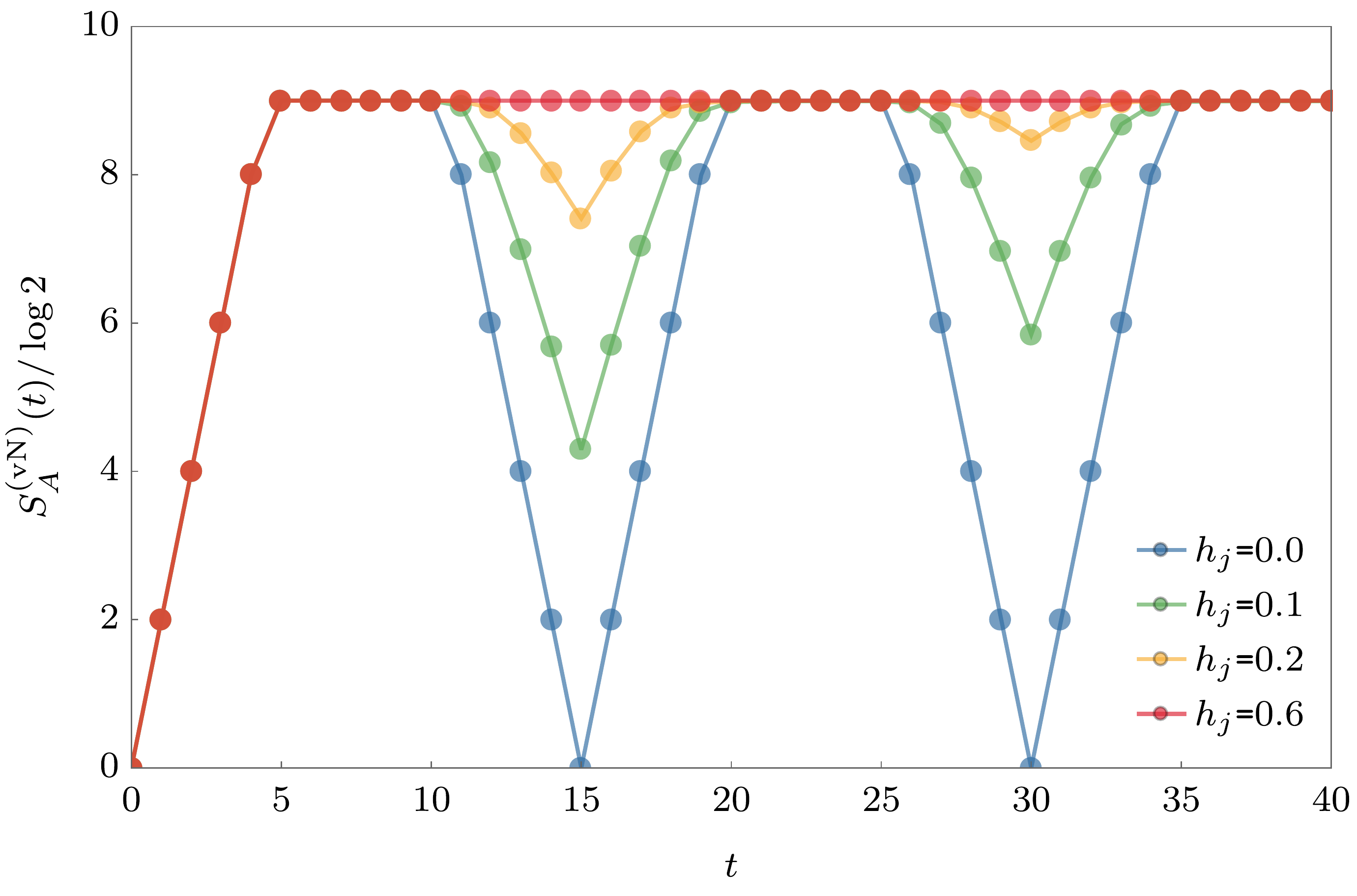}
\caption{The von Neumann entropy (\emph{cf}.~\eqref{eq:vonNeumann}) for a subsystem of $N=9$ spins in a kicked Ising system of $L=30$ evolving from the separating state with $\theta_i=\pi/2$ and $\phi_i=0$ for different values of the longitudinal magnetic field.}
\label{fig:Revivals}
\end{figure*}

A natural question is what happens to the finite-size oscillations when the integrability is weakly broken. This is investigated in Fig.~\ref{fig:Revivals}, which compares the behaviour of the von Neumann entropy (\emph{cf}.~\eqref{eq:vonNeumann}) for increasing values of the (homogeneous) longitudinal magnetic field. We see that the finite-size oscillations become damped and disappear at large enough times. This can be interpreted as a sign of the decay of the quasiparticles. Consistently, the decay speed increases with the magnitude of the longitudinal magnetic field. Moreover, for any fixed $\boldsymbol h\neq0$ the peaks are observed to decay when the volume of the system increases.

\begin{figure*}[t]
\includegraphics[width=0.7\textwidth]{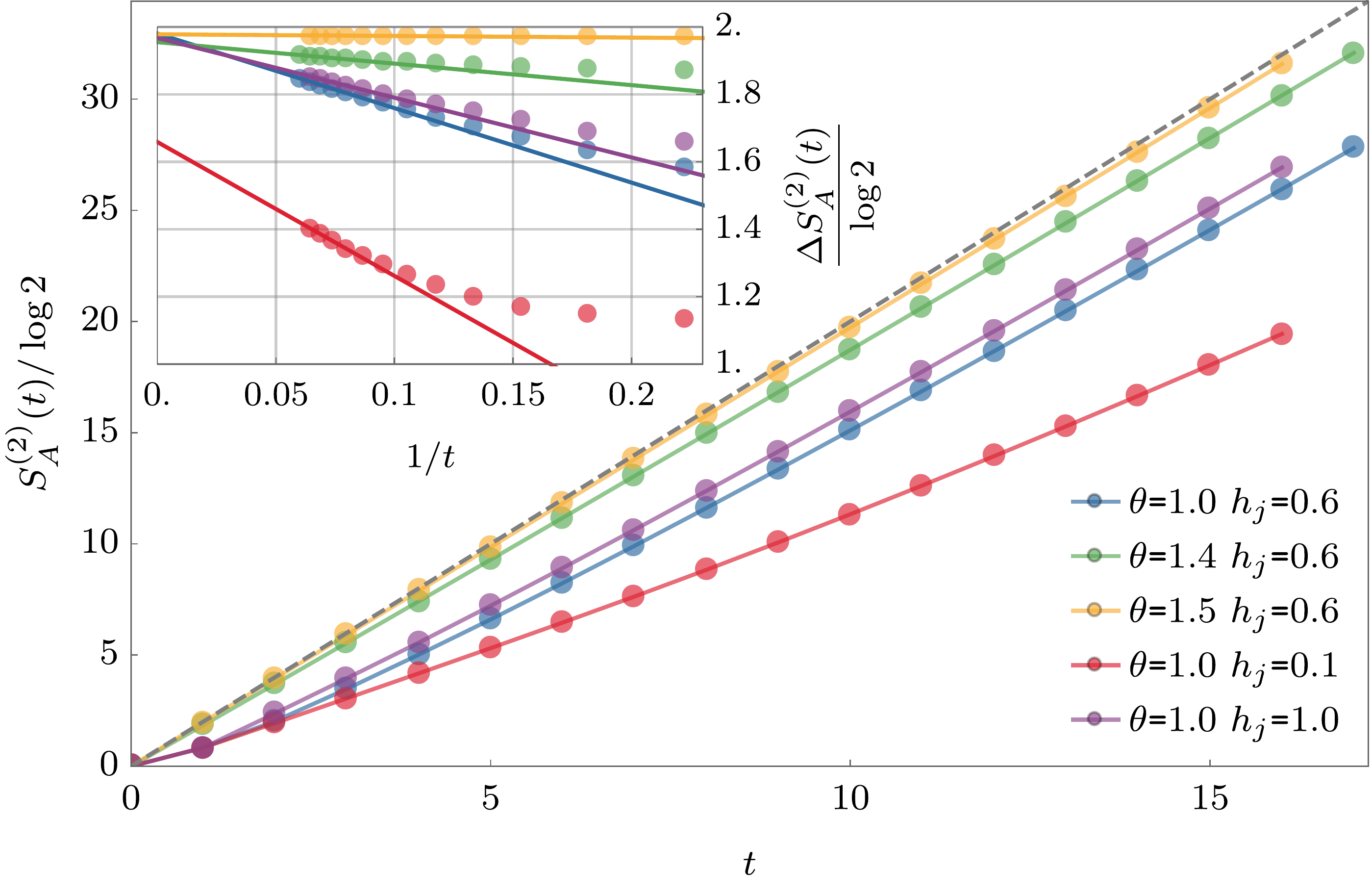}
\caption{The second R\'enyi entropy for a kicked Ising system evolving from ``tilted'' initial states \eqref{eq:state0}  in the thermodynamic limit. The coloured lines correspond to non-separating initial states (we took $\phi=\theta$, with $\theta$ and $h$ specified in the legend) and have been determined numerically evaluating \eqref{eq:TLentropyhomonumNinftyA}, while the grey dashed lines reports, for comparison, the result from separating states. The inset shows the ``instantaneous" slope $\Delta S_A^{(2)}(t)$ (\emph{cf.}~\eqref{eq:instantaneaousslope}), as a function of $1/t$. The points are computed by evaluating \eqref{eq:TLentropyhomonumNinftyA}, while the lines are a linear fit of the last $2$ points. }
\label{fig:DualMethodDiffStates}
\end{figure*}

\begin{figure*}[t]
\includegraphics[width=0.7\textwidth]{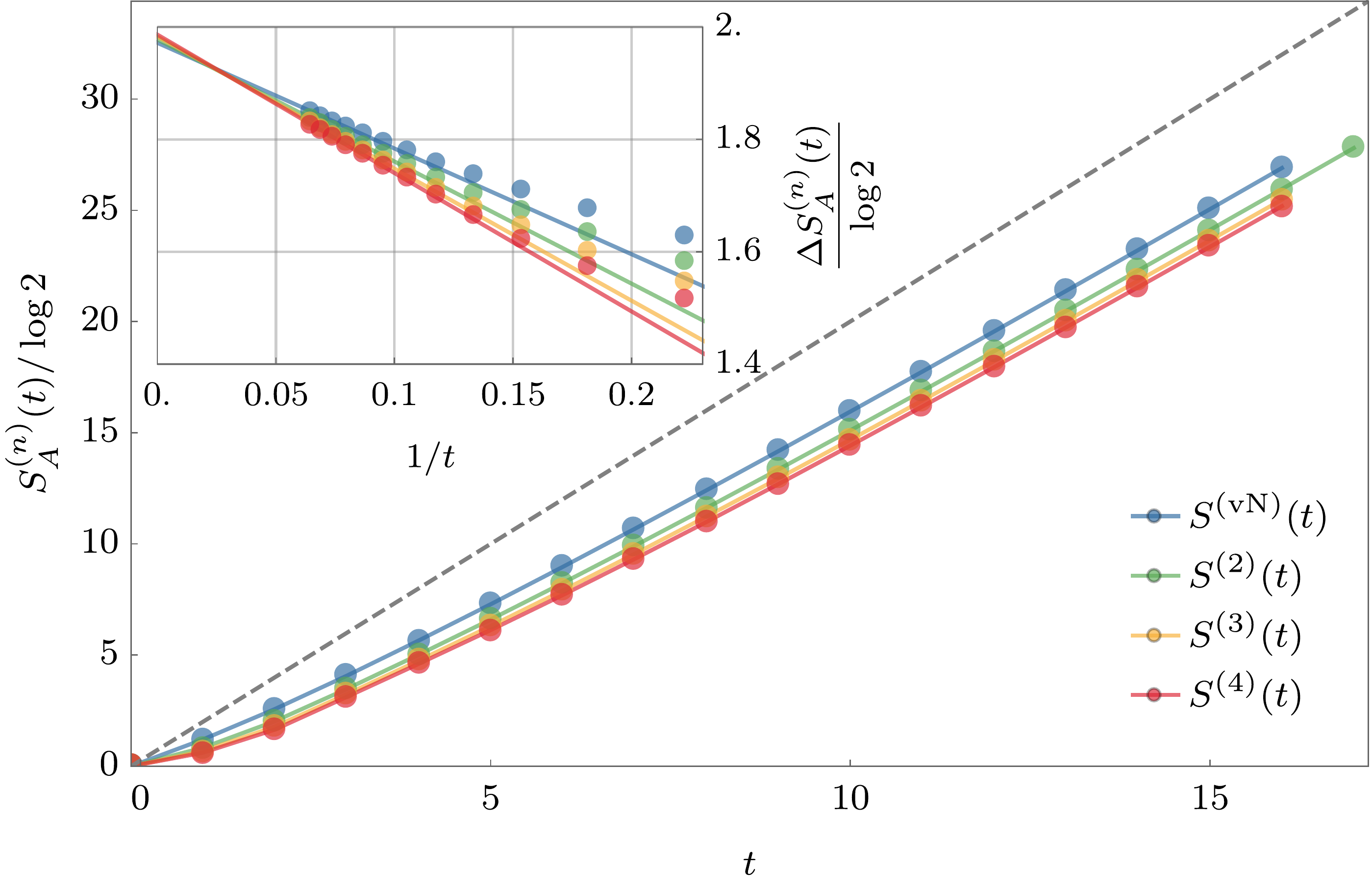}
\caption{Time evolution of different R\'enyi entropies in a kicked Ising system with longitudinal magnetic field $h=0.6$ in the thermodynamic limit. The initial state is of the form \eqref{eq:state0} with $\theta=\phi=1$. The coloured lines correspond to Renyi indices $n=1,2,3,4$ and have been determined numerically evaluating \eqref{eq:TLentropyhomonumNinftyA}, while the grey dashed lines reports, for comparison, the result from separating states. The inset shows the ``instantaneous" slope $\Delta S_A^{(n)}(t)$ (\emph{cf.}~\eqref{eq:instantaneaousslope}), as a function of $1/t$. The points are computed by evaluating \eqref{eq:TLentropyhomonumNinftyA}, while the lines are linear extrapolations from the last $2$ data points.}
\label{fig:DualMethodDiffS}
\end{figure*}

Finally, we note that Fig.~\ref{fig:DiffStates} also contains some information on the speed of entanglement growth. Indeed, we see that the time evolution of the entropies depends (although weakly for ${\boldsymbol h\neq\boldsymbol 0}$) on the configuration of magnetic fields, indicating that the feature (ii) is lost at short times. It is, however, very hard to make any definitive statement based on Fig.~\ref{fig:DiffStates}. The direct numerical approach allows us to access the linear growth regime only for very short times and it is impossible to exclude that (ii) remains as an asymptotic feature of the entanglement dynamics. A similar argument holds regarding the entanglement spectrum.  

A useful way to circumvent this problem is offered by the expression for the R\'enyi entropies resulting from the duality mapping, namely Eq.~\eqref{eq:entropyduality} (a similar duality-based approach has also been proposed in the context of tensor networks~\cite{BHVC}). This becomes particularly convenient in the translational invariant case 
\be
h_j=h,\qquad\theta_j=\theta,\qquad \phi_j=\phi,\qquad \forall j\,.
\ee
Indeed, in this case one can use the general constraints \eqref{eq:physicalrequest} to take analytically the limit of infinite $L$ and $N$ and focus on the growth regime. Specifically, for $n=2,3,\ldots$, we find 
\be
\lim_{N\to\infty}\lim_{L\rightarrow \infty} S^{(n)}_A(t)=\frac{2}{1-n}\log\!\left | \frac{\braket{M_{\rm L}| \mathbb P |M_{\rm R}}}{\braket{M_{\rm L}|M_{\rm R}}}\right|\!, 
\label{eq:TLentropyhomonumNinfty}
\ee
where $\bra{M_{\rm L}}$ and $\ket{M_{\rm R}}$ are respectively the left and right eigenstates of $\mathbb T_{\theta,\phi}[h]$ corresponding to the eigenvalue $1$. The constraints~\eqref{eq:physicalrequest} imply that these vectors exist and are unique. Note that the simplification \eqref{eq:TLentropyhomonumNinfty} cannot be generically performed in the inhomogeneous case, since transfer matrices with different $h,\theta$ and $\phi$ have different left and right eigenvectors. 

The numerical evaluation of \eqref{eq:TLentropyhomonumNinfty} is achieved in two steps. First one has to determine the left and right eigenvectors and then to evaluate the matrix element. Finding eigenvectors is particularly convenient due to the tensor product structure of the transfer matrix (\emph{cf}. \eqref{eq:dualtranfermatrixentropy}). Indeed, we can search eigenvectors of the form
\be
\ket{M_{\rm R}}= \bigotimes_{\nu=1}^{n}\ket{A_{\rm R}},\qquad \bra{M_{\rm L}}= \bigotimes_{\nu=1}^{n}\bra{A_{{\rm L}}}\!,
\label{eq:productform}
\ee
where ${\ket{A_{\rm R}},\ket{A_{\rm L}}\in {\mathcal H_{t}\otimes\mathcal H_{t}}}$. In the notation of Fig.~\ref{fig:tensorproductspace}, this means that we can effectively work in the Hilbert space of the $\nu$-th copy in both the positive-time and negative-time spaces (corresponding to the $\nu$-th column of Fig.~\ref{fig:tensorproductspace}).
The eigenvectors are efficiently determined by means of a simple ``power method": one starts from a random vector and finds $\ket{A_{\rm R}}$  by repeated application of $\mathbb T_{\theta,\phi}[h]$. The left eigenvector $\bra{A_{\rm L}}$ is then determined by using $ \ket{A_{\rm L}} = e^{-i\frac{\pi}{8}M^x_1} \otimes e^{i\frac{\pi}{8}M^x_1} \ket{A_{\rm R}}$. Proceeding in this way the eigenvector can be determined in ${\cal O}(t m 2^{2t})$ operations, where $m$ is the number of iterations of the power method~\cite{notepowermethod}.

The form \eqref{eq:productform} is also convenient for evaluating the matrix element in \eqref{eq:TLentropyhomonumNinfty}. Indeed, after a straightforward calculation we find 
\be
\lim_{N\to\infty}\lim_{L\rightarrow \infty} S^{(n)}_A(t)=\frac{2}{1-n}\log\frac{{\rm tr}[(A_{\rm R}^\dag A^{\phantom \dag}_{\rm L})^n]}{\bigl[{\rm tr}(A_{\rm R}^\dag A^{\phantom \dag}_{\rm L})\bigr]^n}
\label{eq:TLentropyhomonumNinftyA}
\ee
where $A_{\rm L,\rm R}$ are the $2^t\times2^t$ matrices corresponding to the vectors $\bra{A_{\rm R,L}}$ through the vector-to-operator mapping \eqref{eq:statetoop} (performed for ${n=1}$). Note that for transverse separating initial states we have 
\be
A_{\rm R}=A_{\rm L}\propto\1\,,
\ee
but for generic initial states these matrices become non-trivial. The r.h.s. of Eq.~\eqref{eq:TLentropyhomonumNinftyA} can be numerically evaluated for integer $n \ge 2$, whereas R\'enyi entropies with more general index $\alpha>0$ can be found by analytically continuing \eqref{eq:TLentropyhomonumNinftyA} in $n$ and diagonalising $A_{\rm R}^\dag A^{\phantom \dag}_{\rm L}$ numerically to compute the powers. Evaluating ~\eqref{eq:TLentropyhomonumNinftyA} has complexity $\propto 2^{3t}$ and is the bottleneck of the numerical procedure, meaning that we were able to reach up to ${t_{\rm max}=17}$.

We stress that, since the constraints~\eqref{eq:physicalrequest} hold also away from the self dual points, this procedure can be used to study the entanglement spreading in the \emph{entire parameter space} of the kicked Ising model. Note that close enough to the self-dual points (and to separating initial states) an analytical, perturbative, analysis is also possible. These aspects, however, go beyond the scope of the present manuscript and will be investigated the course of future research. Here we focus on the self-dual points \eqref{eq:selfdual} and use this duality-based numerical approach to effectively investigate the fate of the features (ii) and (iii) when the system is initialised in a generic state \eqref{eq:state0}. 

Representative examples of our numerical results are reported in Figs.~\ref{fig:DualMethodDiffStates} and \ref{fig:DualMethodDiffS}. We see that, consistently with the results in Fig.~\ref{fig:DiffStates}, at short times both (ii) and (iii) are violated. The entropies grow in an approximately linear fashion but the the slope appears to depend on the initial state and on the longitudinal magnetic field $h$ (see Fig.~\ref{fig:DualMethodDiffStates}). Moreover, different R\'enyi entropies have different slopes (see  Fig.~\ref{fig:DualMethodDiffS}). Crucially, however, a more refined analysis suggests that these deviations \emph{vanish for large times}. To show this we proceed as follows. First we introduce the the ``instantaneous" slopes
\be
\Delta S_A^{(\alpha)}(t-1/2)\equiv S_A^{(\alpha)}(t)- S_A^{(\alpha)}(t-1).
\label{eq:instantaneaousslope}
\ee
Plotting the instantaneous slopes as functions of $1/t$ we see that they become approximately linear at large enough times. We then perform a linear fit and extrapolate the result to $t=\infty$. In the integrable case this procedure gives results consistent with the quasiparticle picture prediction (\emph{cf}. Eq.~\eqref{eq:scalingfree}), namely
\be
\Delta S_A^{(\alpha)}(\infty)\bigr |_{h=0}=2 S^{(\alpha)}_{\theta,\phi} \leq 2 \log 2\,. 
\label{eq:integrableslope}
\ee
Instead, in the non-integrable case the results are consistent with 
\be
\Delta S_A^{(\alpha)}(\infty)\bigr |_{h\neq0}=2\log 2.
\label{eq:nonintegrableslope}
\ee
This is observed for any initial state~\eqref{eq:state0}, for any non-vanishing longitudinal magnetic field, and for any R\'enyi index $\alpha$, as demonstrated in the insets of Figs.~\ref{fig:DualMethodDiffStates} and \ref{fig:DualMethodDiffS}. The only seemingly exceptional cases are observed when the system is very close to the integrable point (see, \emph{e.g.}, the red curve in the inset of Fig.~\ref{fig:DualMethodDiffStates}). This is, however, straightforwardly explained as a prethermalization effect~\cite{MK:pret,BEGR:pret,Pretrev}. For small enough longitudinal fields there is an initial transient, before the quasiparticles decay, in which the observables follow the integrable predictions. After this transient, however, the entropies are expected to follow the non-integrable curves, reaching the asymptotic value \eqref{eq:nonintegrableslope} for the slope. For instance, this is consistent with the behaviour of the red curve in the inset of Fig.~\ref{fig:DualMethodDiffStates}. Interestingly, since $S^{(\alpha)}_{\theta,\phi} = \log 2$ for separating states, this effect is not observed in our exact result~\eqref{eq:finalresult}. From the physical point of view \eqref{eq:nonintegrableslope} is very natural. Since the system is ergodic the initial state dependence is washed away at large enough times, and the entropies behave as if they would be evolving from separating states. As expected, this does not happen in the integrable case.

In conclusion, the numerical results presented in this section support in the following general picture. The evolution of the entanglement entropies from a generic initial state~\eqref{eq:state0} in the thermodynamic limit differs from that from separating states (\emph{cf}.~\eqref{eq:finalresult}), and depends explicitly on the initial state, the longitudinal magnetic fields, and the R\'enyi numbers. In the scaling limit $t,N\to\infty$, however, all these dependences are washed away: the entropies collapse to the prediction \eqref{eq:finalresult} if the system is non-integrable and to the thermodynamic limit of \eqref{eq:scalingfree} if the system is integrable. 
In other words, our exact result~\eqref{eq:finalresult} serves as an asymptotic description of the entanglement spreading in the non-integrable kicked Ising model.
This picture is also supported by the numerical results of Ref.~\cite{PL:kickedIsing}, which found that the evolution of the von Neumann entropy averaged over all separable initial states, is consistent with~\eqref{eq:finalresult}. 
Finally, we remark that this section also demonstrated that extracting universal information on the behaviour of the entanglement entropies from the numerics is extremely hard, especially in the ergodic case where the entanglement growth is exceptionally fast. This highlights even more the practical importance of our exact result~\eqref{eq:finalresult}.

%%%%%%%%%%%%
\section{Conclusions}%%
%%%%%%%%%%%%

\label{sec:conclusions}

We have developed a constructive and mathematically rigorous approach for computing the dynamics of bipartite entanglement in a class of ``{maximally scrambling}", locally interacting, chaotic spin chains. Specifically, we considered 
the so called ``self-dual" kicked Ising spin chains, where the integrability is broken by switching on an external longitudinal magnetic field. We prepared the system in class of ground states of simple local Hamiltonians and determined exactly the dynamics of all R\'enyi entropies of finite blocks of spins of arbitrary size. The results presented are non-perturbative, no kind of averaging is involved, and, most importantly, they hold in the presence of longitudinal magnetic fields with arbitrary spatial dependence. It is remarkable that such an explicit exact calculation can be performed for a specific non-integrable many-body system.  

Our result shows that in the thermodynamic limit the R\'enyi entropies of finite blocks of spins are independent of the longitudinal magnetic field at all times. Moreover, they obey universal scaling laws that can be predicted both by means of the quasiparticle picture of Ref.~\cite{CC}, put forward for integrable models, and of the minimal membrane picture of Ref.~\cite{RandomCircuitsEnt}, propounded for generic systems. 

Using our novel rigorous approach, we also developed a numerical procedure for studying the entanglement spreading from generic product initial states. A thorough numerical analysis suggests that, away from the integrable point, our exact result continues to describe the entanglement spreading at the leading order in time. On the contrary, in the integrable case the entanglement production is generically  renormalised by an initial-state-dependent multiplicative coefficient. Further qualitative differences between the integrable and the non-integrable case emerge for finite systems. In particular, we showed numerically that there are recurrences in the integrable case, which are absent in the non-integrable one. We stress that these differences are correctly accounted for by the quasiparticle and minimal membrane pictures, which disagree for finite sizes.

Our analytical method can be used to highlight qualitative differences in the entanglement spreading of integrable and non-integrable systems directly in the thermodynamic limit. To do that one could follow Refs.~\cite{ABGH:CFT, LM:CFT} and consider the bipartite entanglement of disjoint blocks. Our preliminary results suggest that the scaling forms produced in the two cases are indeed different and respectively agree with the predictions of quasiparticle and membrane pictures. Another possible direction is to perturb the kicked Ising spin chains away from the ``self-dual" points, where the predictions of the two pictures disagree also for the entanglement of a single block. This could be tested within our approach by using  perturbation theory.

More generally, we expect that our method would allow for explicit calculations similar to the ones presented also for other measures of correlations and dynamical complexity, such as operator space entanglement entropy and out-of-time order correlators.

Finally, we believe that the remarkable algebraic structure unveiled in this work paves the way for the determination of a new class of exactly solvable, maximally chaotic models. Elements of this class can serve as minimal models for characterising the non-equilibrium dynamics in generic systems.

\bigskip

\section*{Acknowledgements}

We thank Adam Nahum for very valuable comments on the manuscript and Pasquale Calabrese for stimulating discussions. B.B. and T.P. acknowledge the hospitality of the Erwin Schr\" odinger Institute (ESI), Vienna, where this project has been conceived. The work has been supported by Advanced Grant of European Research Council (ERC), No. 694544 -- OMNES and program P1-0402 of Slovenian Research Agency.

\begin{widetext}
\appendix

\section{Duality of traces}
\label{app:duality}

Here we explicitly demonstrate the duality relation (\ref{eq:dual}). Writing ${\rm tr}\left[\left(U_{\rm KI}[\boldsymbol h]\right)^t\right]$ in the computational basis $\mathcal B_L$ (\emph{cf}.~\eqref{eq:computationalbasis}) we have 
\begin{align}
{\rm tr}\left[\left(U_{\rm KI}[\boldsymbol h]\right)^t\right] &= \sum_{\{\boldsymbol s_{\tau}\}} \bra{\boldsymbol s_{1}}U_{\rm KI}[\boldsymbol h]\ket{\boldsymbol s_{t}}\braket{\boldsymbol s_{t}|U_{\rm KI}[\boldsymbol h]|\boldsymbol s_{t-1}}\cdots\braket{\boldsymbol s_{2}|U_{\rm KI}[\boldsymbol h]|\boldsymbol s_1} \notag\\
&= \left(\frac{\sin 2b}{2i}\right)^{\frac{Lt}{2}} \!\!\!\!\sum_{\{ s_{\tau,j}\}}\left\{ \exp\!\!\left[- i \tilde J \sum_{j=1}^L s_{1,j}s_{t,j}- i J \sum_{j=1}^L s_{t,j} s_{t,j+1}-i \sum_{j=1}^L h_j s_{t,j}\right]\right.\notag\\
&\qquad\qquad\quad\;\;\;\, \times \exp\!\!\left[- i \tilde J \sum_{j=1}^L s_{t,j}s_{t-1,j}- i J  \sum_{j=1}^L s_{t-1,j} s_{t-1,j+1}-i \sum_{j=1}^L h_j s_{t-1,j}\right]\notag\\
&\qquad\qquad\qquad\qquad\qquad\qquad\qquad\qquad\qquad\qquad\vdots\notag\\
&\left.\qquad\qquad\quad\;\;\;\, \times \exp\!\!\left[ - i \tilde J \sum_{j=1}^L s_{2,j} s_{1,j}- i J \sum_{j=1}^L s_{1,j} s_{1,j+1}-i \sum_{j=1}^L h_j s_{1,j}\right]\right\}\,.
\label{eq:appdualitystep1}
\end{align}
Here $s_{\tau,L+j}\equiv s_{\tau,j}$ and in the second step we used the identity
\be
\braket{s|e^{-i b \sigma^x}|r}= \sqrt{\frac{\sin 2b}{2i}} \exp\left[ - i \tilde J  s r\right],\qquad s,r\in\{\pm1\}\,,
\label{eq:identityapp}
\ee
where
\be
\tilde J= -\frac{\pi}{4}-\frac{i}{2}\log\tan b\,.
\ee

This expression can be thought of as the partition function of a two-dimensional Ising model with complex couplings on a $L\times t$ periodic lattice. In other words, the r.h.s. of \eqref{eq:appdualitystep1} is proportional to the partition function of a classical statistical mechanical model with configuration energy given by 
\be
\mathcal E[\{s_{\tau,j}\},\boldsymbol{h}] = - \sum_{\tau=1}^t  \sum_{j=1}^{L} (iJ s_{\tau,j} s_{\tau,j+1} + i\tilde J  s_{\tau,j} s_{\tau+1,j} + ih_j s_{\tau,j})\,.
\ee

Reorganising the sum on the r.h.s. of \eqref{eq:appdualitystep1} we also have 
\begin{align}
{\rm tr}\left[\left(U_{\rm KI}[\boldsymbol h]\right)^t\right] &=  \left(\frac{\sin 2b}{2i}\right)^{\frac{Lt}{2}} \!\!\!\!\sum_{\{ s_{\tau,j}\}}\left\{ \exp\!\!\left[- i\tilde J\sum_{\tau=1}^{t} s_{\tau,1}s_{\tau+1,1}- i J  \sum_{\tau=1}^t s_{\tau,1} s_{\tau,L}-i \sum_{\tau=1}^t h_1 s_{\tau,1}\right]\right.\notag\\
&\left.\qquad\qquad\quad\quad \times
\exp\!\!\left[- i\tilde J \sum_{\tau=1}^{t} s_{\tau,2}s_{\tau+1,2}- i J \sum_{\tau=1}^t s_{\tau,1} s_{\tau,2}-i \sum_{\tau=1}^t h_{2} s_{\tau,2}\right]\right.\notag\\
&\qquad\qquad\qquad\qquad\qquad\qquad\qquad\qquad\qquad\qquad\vdots\notag\\
&\left.\qquad\qquad\quad\quad \times\exp\!\!\left[- i\tilde J \sum_{\tau=1}^{t} s_{\tau,L}s_{\tau+1,L}- i J \sum_{\tau=1}^t s_{\tau,L} s_{\tau,L-1}-i \sum_{\tau=1}^t h_L s_{\tau,L}\right]\right\}\,,
\end{align}
where we defined $s_{t+\tau,j}\equiv s_{\tau,j}$. Using again the identity \eqref{eq:identityapp} we finally find 
\be
{\rm tr}\left[\left(U_{\rm KI}[\boldsymbol h]\right)^t\right] = {\rm tr} \left(  \tilde{U}_{\rm KI}[h_1 \tilde{\boldsymbol 1}]\cdots \tilde{U}_{\rm KI}[h_L \tilde{\boldsymbol 1}]\right)\,,
\ee 
where ``tilded''  bold symbols denote vectors of $t$ components and we introduced the dual transfer matrix
\be
\tilde U_{\rm KI}[\tilde{\boldsymbol h}]= e^{-i \tilde H_{\rm K}}e^{-i \tilde H_{\rm I}[\tilde{\boldsymbol h}]}\,,
\ee
with 
\be
\tilde H_{\rm I}[\tilde{\boldsymbol h}]\equiv\! \tilde J \sum_{j=1}^{t} \sigma^{z}_j \sigma^z_{j+1}+ \sum_{j=1}^{t} h_j \sigma^z_{j}\,,\qquad\qquad\tilde H_{\rm K}\equiv \tilde b \sum_{j=1}^{t}  \sigma^x_j.
\ee

\section{Simplified transfer matrix for longitudinal separating states}
\label{app:longstates}

When the initial state is in the class $\cal L$ (\emph{cf}.~\eqref{eq:classesL}), namely when 
\be
\theta_{j}=\bar\theta_j\equiv(1+s_{j}) \pi/2,\qquad s_j\in\{-1,1\},\qquad j \in\{1,2,\ldots,L\},
\ee 
the form \eqref{eq:entropyduality} can be simplified by effectively reducing the dimension of the space where the trace acts. To see this we note that in this case $\mathbb B^{z}_{\nu, 1}[\theta]$ becomes proportional to a projector:
\be
\mathbb B^{z}_{\nu, 1}[(1+s_j) \pi/2] =  2 P_{\nu,1}^{z,s_{j}}\otimes P_{\nu,1}^{z,s_{j}}\,,
\ee
so that we have 
\begin{align}
&{\rm tr}\left[ \left(\prod_{j=1}^N \mathbb T_{\bar \theta_j,\phi_j}[h_j]\right)\mathbb P\,  \left (\prod_{j=N+1}^L \mathbb T_{\bar \theta_{j},\phi_{j}}[h_{j}]\right)\mathbb P^{\dag}\right]\notag\\
&= 2^{L n} {\rm tr}\left[\prod_{j=1}^L\prod_{\nu=1}^n P_{\nu,1}^{z,s_j}e^{i\frac{\pi}{4}  \sigma^x_{\nu,1}}\otimes P_{\nu,1}^{z,s_j}e^{-i\frac{\pi}{4}  \sigma^x_{\nu,1}}\right] {\rm tr}\left[\left(\prod_{j=1}^N \bar{\mathbb T}_{\frac{\pi}{2},\bar \theta_j-\frac{\pi}{2}}[h_j]\right)\bar{\mathbb P}\,  \left (\prod_{j=N+1}^L \bar{\mathbb T}_{\frac{\pi}{2},\bar \theta_j-\frac{\pi}{2}}[h_{j}]\right)\bar{\mathbb P}^{\dag}\right], \label{eq:traces}
\end{align}
where we introduced 
\begin{align}
\bar{\mathbb P} &\equiv  \1\otimes\prod_{\nu=1}^{n}\prod_{\tau=2}^{t}  P_{(\nu,\tau),(\nu-1,\tau)}\,, \\
\bar{\mathbb T}_{\theta, \phi}[h]&\equiv\mathbb B^{z}_{\nu, 2}[\theta] \cdot \mathbb G^{z}_{\nu, t}\cdot \bar{\mathbb U}_{\phi}[h].
\end{align}
Here the matrix $\bar{\mathbb U}_{\phi}^{(\nu)}[h]$ is defined as 
\be
\bar{\mathbb U}_{\phi}^{(\nu)}[h]\equiv  
\left(\bar{U}_{\nu,\phi}\otimes \bar{U}_{\nu,\phi}^* \right)\cdot\left(e^{-i h  \bar M^z_{\nu}} \otimes e^{i h \bar M^z_{\nu}}\right)\cdot\left(e^{i\frac{\pi}{4}  \bar M^x_{\nu}} \otimes e^{-i\frac{\pi}{4} \bar M^x_{\nu}}\right),
\ee
and the barred operators read as 
\be
\bar {U}_{\nu,\phi}\equiv \exp\left[-\frac{i\pi}{4} \sum_{\tau=2}^{t-1} \sigma^{z}_{\nu, \tau}  \sigma^z_{\nu, \tau+1}-i\frac{\phi}{2}\sigma^{z}_{\nu,2}\right],\qquad\qquad \bar {M}^a_{\nu}\equiv  \sum_{\tau=2}^{t} \sigma^{a}_{\nu, \tau}\,.
\label{eq:Ubar}
\ee
So the have the same form as \eqref{eq:defU} and \eqref{eq:defM} but at fixed $\nu$ they act non-trivially only in the space $\mathcal H_{t-1}$ composed of the last $t-1$ sites of $\mathcal H_t$. In other words $\bar{\mathbb T}_{\theta, s}[h]$ has the same form as ${\mathbb T}_{\theta, s}[h]$ but acts on $\mathcal H_{t-1}^{\otimes 2n}$ instead of $\mathcal H_{t}^{\otimes 2n}$. We stress that the trace operations in expression (\ref{eq:traces}) and below are taken in the subspaces where the operators act nontrivially, for example, for the barred operators in 
${\cal H}^{\otimes 2n}_{t-1}\cong\mathcal {\cal H}_{2n(t-1)}$. 
Noting 
\be
2^{L n} {\rm tr}\left[\prod_{j=1}^L\prod_{\nu=1}^n P_{\nu,1}^{z,s_j}e^{i\frac{\pi}{4}  \sigma^x_{\nu,1}}\otimes P_{\nu,1}^{z,s_j}e^{-i\frac{\pi}{4}  \sigma^x_{\nu,1}}\right]=2^{L n} \left|{\rm tr}\left[\prod_{j=1}^L \left[P_{1,1}^{z,s_j}e^{i\frac{\pi}{4}  \sigma^x_{1,1}}\right]\right]\right|^{2n}\!\!\!\!\!=1\qquad\qquad \forall\, s_j\in\{-1,+1\}\,,
\ee
we finally find 
\be
S^{(n)}_A(t)=\frac{1}{1-n}\log {\rm tr}\left[\left(\prod_{j=1}^N \bar{\mathbb T}_{\frac{\pi}{2},\bar \theta_j-\frac{\pi}{2}}[h_j]\right)\bar{\mathbb P}\,  \left (\prod_{j=N+1}^L \bar{\mathbb T}_{\frac{\pi}{2},\bar \theta_j-\frac{\pi}{2}}[h_{j}]\right) \bar{\mathbb P}^{\dag}\right].
\label{eq:entropyduality2}
\ee

Therefore we see that in this case the entropies are given by an expression of the form \eqref{eq:entropyduality}, with $\theta_j=\tfrac{\pi}{2}$ and $\phi_j=\tfrac{\pi}{2}s_j$,  but with matrices acting on $\mathcal H_{t-1}^{\otimes 2n}$ instead of $\mathcal H_{t}^{\otimes 2n}$. Note that for $\theta_j=\bar \theta_j$ the states \eqref{eq:state0} do not depend on $\phi_i$ and this independence is correctly reflected in Eq.~\eqref{eq:entropyduality2}.

\section{Proof of Property~\ref{prop:transverse}}
\label{app:proofsp1p2}

In this appendix we provide the proof of Property~\ref{prop:transverse}.  
\begin{proof}
For each state $\bra{A}$ we have 
\begin{align}
&\braket{A|\mathbb T^{\phantom \dag}_{\theta,\phi}[h]\mathbb T^\dag_{\theta,\phi}[h]|A}= \braket{A|\prod_{\nu=1}^n\mathbb B^{z}_{\nu, 1}[\theta]\prod_{\nu=1}^n \mathbb G^{z}_{\nu, t}\prod_{\nu=1}^n\mathbb B^{z}_{\nu, 1}[\theta]|A}\leq \braket{A|\prod_{\nu=1}^n\mathbb B^{z}_{\nu, 1}[\theta]^2|A},
\label{eq:boundTT}
\end{align}
where we used that $\mathbb G^{z}_{\nu, t}$ is a projector, so its expectation value on a normalised state is smaller or equal to one. Expanding $\mathbb B^{z}_{\nu, 1}[\theta]^2$ we then have  
\begin{align}
\braket{A|\mathbb T^{\phantom \dag}_{\theta,\phi}[h]\mathbb T^\dag_{\theta,\phi}[h]|A}\leq 4^n \bra{A}\prod_{\nu=1}^n &\left(\cos^4(\theta/2) P_{\nu,1}^{z,+}\otimes P_{\nu,1}^{z,+} + \sin^2(\theta/2) \cos^2(\theta/2) P_{\nu,1}^{z,-}\otimes P_{\nu,1}^{z,+}\right. \notag\\
&\quad+ \left.\sin^2(\theta/2) \cos^2(\theta/2) P_{\nu,1}^{z,+}\otimes P_{\nu,1}^{z,-} + \sin^4(\theta/2)  P_{\nu,1}^{z,-}\otimes  P_{\nu,1}^{z,-} \right)\ket{A}.
\end{align}
Since $P_{\nu,1}^{z,\pm}\otimes P_{\nu,1}^{z,\pm}$ are orthogonal projectors we have 
\be
\braket{A|\mathbb T^{\phantom \dag}_{\theta,\phi}[h]\mathbb T^\dag_{\theta,\phi}[h]|A}\leq 4^n \max(\sin^{4n}(\theta/2),\cos^{4n}(\theta/2))\,.
\label{eq:boundw}
\ee
In particular, choosing $\bra{A}$ to be the left eigenstate of $\mathbb T_{\theta,\phi}[h]$ corresponding to the eigenvalue $\lambda$ we have
\be
|\lambda|\leq 2^n \max(\sin^{2n}(\theta/2),\cos^{2n}(\theta/2))=(1+|\cos\theta|)^n= \lambda_{\rm max}\,,
\ee
which proves the first part of the claim. 

To prove the point $(ii)$ $a.$ we proceed by \emph{reductio ad absurdum}. Suppose that the Jordan block of $\lambda$ is non-trivial: let $\bra{A}$ be the eigenvector associated to $\lambda$ and let $\bra{B}$ be the first generalised eigenvector. As it is always possible, we choose $\bra{B}$ to be normalised and orthogonal to $\bra{A}$ (which is also normalised). We then have 
\be
\bra{B} \mathbb T_{\theta,\phi}[h] = \lambda \bra{B} +x \bra{A}\qquad\qquad\qquad\qquad\qquad x\neq0\,.
\ee
This implies 
\be
\braket{B|\mathbb T^{\phantom \dag}_{\theta,\phi}[h] \mathbb T^\dag_{\theta,\phi}[h]|B} =|\lambda_{\rm max}|^2 +|x|^2\,,
\ee
which is impossible because it contradicts \eqref{eq:boundw}. Point $(ii)$ $b.$ follows by noting that in order to have the equality sign in \eqref{eq:boundw} we must have 
\begin{align}
\bra{A}\prod_{\nu=1}^n\mathbb B^{z}_{\nu, 1}[\theta]&=\lambda_{\rm max}\bra{A},\\
\bra{A}\prod_{\nu=1}^n\mathbb G^{z}_{\nu, t}&=\bra{A}\,.
\end{align}
Using now that $\bra{A}$ is a left eigenvector of $\mathbb T_{\theta,\phi}[h]$ we have \eqref{eq:eigTT}. This concludes the proof. 

\end{proof}

\section{Simplified commutation relations}
\label{app:proofsp3p4}

In this appendix we prove the following property. 
\begin{property}
\label{prop:basicproperty}
The commutation relations \eqref{eq:condAo1}--\eqref{eq:condAo2} imply
\be
 A \sigma^a_{\nu,\tau} = \sigma^a_{\nu,\tau} A\,, \qquad\qquad\qquad\qquad\qquad\forall\;\; a\in\{x,y,z\},\;\; \tau\in\{1,\ldots, t\},\;\; \nu\in\{1,\ldots,n\}\,.
\ee
\end{property}

\begin{proof}
First of all we note that multiplying \eqref{eq:condAo2} on the left and on the right by $e^{-i\frac{\pi}{4}  M_\nu^x}\, e^{i h  M_\nu^{z}}{U}^\dag_{\nu,\phi}$ we have 
\be
A\,e^{-i\frac{\pi}{4}  M_\nu^x}\, e^{i h  M_\nu^{z}}{U}^\dag_{\nu,\phi}= e^{i \alpha_\nu} e^{-i\frac{\pi}{4}  M_\nu^x}\, e^{i h  M_\nu^{z}}{U}^\dag_{\nu,\phi}\, A\,,\qquad \alpha_{\nu}\in\mathbb R\,,\qquad\forall\,\,\nu\in\{1,\ldots,n\}\,.
\label{eq:condAo3}
\ee
Using the conditions \eqref{eq:condAo1}, \eqref{eq:condAo2}, and \eqref{eq:condAo3} we see that $A$ commutes with 
\be
e^{-i\frac{\pi}{4}  M_\nu^x}\, e^{i h  M_\nu^{z}}{U}^\dag_{\nu,\phi}\,\sigma^z_{\nu, t} {U}_{\nu,\phi}\,  e^{-i h  M_\nu^{z}}\, e^{i\frac{\pi}{4}  M_\nu^x} = e^{-i\frac{\pi}{4}  M_\nu^x}\,\sigma^z_{\nu, t}\,e^{i\frac{\pi}{4}  M_\nu^x} = -\sigma^y_{\nu, t}\,. 
\ee
Indeed we have 
\begin{align}
& A\, e^{-i\frac{\pi}{4}  M_\nu^x}\, e^{i h  M_\nu^{z}}{U}^\dag_{\nu,\phi}\,\sigma^z_{\nu, t} {U}_{\nu,\phi}\,  e^{-i h  M_\nu^{z}}\, e^{i\frac{\pi}{4}  M_\nu^x}=\notag\\
& e^{i \alpha_\nu}  e^{-i\frac{\pi}{4}  M_\nu^x}\, e^{i h  M_\nu^{z}}{U}^\dag_{\nu,\phi}\,A\,\sigma^z_{\nu, t} {U}_{\nu,\phi}\,  e^{-i h  M_\nu^{z}}\, e^{i\frac{\pi}{4}  M_\nu^x}=\notag\\
& e^{i \alpha_\nu} e^{-i\frac{\pi}{4}  M_\nu^x}\, e^{i h  M_\nu^{z}}{U}^\dag_{\nu,\phi}\,\sigma^z_{\nu, t}\,A\, {U}_{\nu,\phi}\,  e^{-i h  M_\nu^{z}}\, e^{i\frac{\pi}{4}  M_\nu^x}=\notag\\
&  e^{-i\frac{\pi}{4}  M_\nu^x}\, e^{i h  M_\nu^{z}}{U}^\dag_{\nu,\phi}\,\sigma^z_{\nu, t} {U}_{\nu,\phi}\,  e^{-i h  M_\nu^{z}}\, e^{i\frac{\pi}{4}  M_\nu^x}\, A\,,\label{eq:identityA}
\end{align}
where in the the first step we used \eqref{eq:condAo3}, in the second \eqref{eq:condAo2}, and in the third \eqref{eq:condAo1}. Using \eqref{eq:condAo1} we then have that $A$ also commutes with 
\be
-i \sigma^y_{\nu, t} \sigma^z_{\nu, t}=\sigma^x_{\nu, t}\,.
\ee
We then have 
\be
\left[ A,\sigma^a_{\nu, t}\right]=0\,,\qquad\qquad\forall\;\;a\in\{x,y,z\}\,,\;\;\nu\in\{1,\ldots,n\}\,.
\label{eq:commuteswithallatsitentm1}
\ee
Using \eqref{eq:commuteswithallatsitentm1}, \eqref{eq:condAo2}, and \eqref{eq:condAo3} we can then conclude the proof by induction. 

We will prove that if 
\be
\left[ A,\sigma^a_{\nu, \tau}\right]=0\,,\qquad\qquad\forall\;\;a\in\{x,y,z\}\,,\;\; \tau \in\{ \bar\tau+1,\bar\tau+2,\ldots,t\}\,,
\label{eq:hypothesis}
\ee
then 
\be
\left[ A,\sigma^a_{\nu, \bar \tau}\right]=0\,,\qquad\qquad\forall \;\; a\in\{x,y,z\}\,,
\label{eq:claim}
\ee
and then proceeding by induction in $\bar \tau=t-1,\ldots,1$. The basis of the induction is given by \eqref{eq:commuteswithallatsitentm1}, so we just need to prove the inductive step. Assuming \eqref{eq:hypothesis} and proceeding as in \eqref{eq:identityA} we can show that $A$ commutes also with 
\begin{align}
&{U}_{\nu,\phi}\,  e^{-i h  M_\nu^{z}}\, e^{i\frac{\pi}{4}  M_\nu^x}\left(\prod_{\tau=
\bar\tau+1}^{t}\sigma^x_{\nu, \tau} \right) e^{-i\frac{\pi}{4}  M_\nu^x}\, e^{i h  M_\nu^{z}}{U}^\dag_{\nu,\phi}\notag\\
&=\left(\prod_{\tau=\bar\tau+1}^{t} -i \sigma^z_{\nu, \tau-1} \sigma^z_{\nu, \tau}  e^{i 2 h \sigma^z_{\nu, \tau}}\sigma^x_{\nu, \tau} \right)= - i^{t-\bar\tau} \sigma^z_{\nu, \bar\tau} \sigma^z_{\nu, t }   
\left(\prod_{j=\bar\tau+1}^{t}e^{i 2 h \sigma^z_{\nu, \tau}}\sigma^x_{\nu, \tau} \right)\,.
\end{align}
The inductive hypothesis \eqref{eq:hypothesis} then implies  
\be
\left[A,\sigma^z_{\nu, \bar\tau}\right]=0\,.
\ee
Reasoning now as in \eqref{eq:identityA} we then have that $A$ also commutes with 
\begin{align}
e^{-i\frac{\pi}{4}  M_\nu^x}\, e^{i h  M_\nu^{z}}{U}^\dag_{\nu,\phi}\, \sigma^z_{\nu, \bar\tau} {U}_{\nu,\phi}\,  e^{-i h  M_\nu^{z}}\, e^{i\frac{\pi}{4}  M_\nu^x} &=e^{-i\frac{\pi}{4}  M_\nu^x}\,  \sigma^z_{\nu, \bar\tau}\,e^{i\frac{\pi}{4}  M_\nu^x}=-\sigma^y_{\nu, \bar\tau}\,,  \\
e^{-i\frac{\pi}{4}  M_\nu^x}\, e^{i h  M_\nu^{z}}{U}^\dag_{\nu,\phi}\, \sigma^z_{\nu, \bar\tau} {U}_{\nu,\phi}\,  e^{-i h  M_\nu^{z}}\, e^{i\frac{\pi}{4}  M_\nu^x}\,\sigma^z_{\nu, \bar\tau} &= e^{-i\frac{\pi}{4}  M_\nu^x}\,  \sigma^z_{\nu, \bar\tau}\,e^{i\frac{\pi}{4}  M_\nu^x}\,\sigma^z_{\nu, \bar\tau}= -i \sigma^x_{\nu, \bar\tau}\,.
\end{align}
So we have 
\be
\left[A,\sigma^a_{\nu, \tau}\right]=0\,,\qquad\qquad a\in\{x,y,z\}\,,\; \tau\in\{1,\ldots,t\}\,,\;\nu\in\{1,\ldots,n\}\,.
\ee
This concludes the proof. 
\end{proof}

\section{Proof of Eqs.~\eqref{eq:stateproperty} and \eqref{eq:statepropertyleft}}
\label{app:proofofstateproperty}

Let us start by proving \eqref{eq:stateproperty}. First we note
\be
\prod_{\nu=1}^{n} O_\nu\otimes O^{*}_\nu \ket{\1}=\ket{\1}\,.\label{eq:statepropertyi2}
\ee
This is explicitly proven as follows
\begin{align}
\prod_{\nu=1}^{n} O_\nu\otimes O^{*}_\nu \ket{\1}&=\frac{1}{2^{n t /2}}\sum_{k,m,m'} \braket{m|\prod_{\nu=1}^{n} O_\nu|k}\bra{m'}\prod_{\nu=1}^{n} O_\nu\ket{k}^* \ket{m}\otimes \ket{m'}^*\notag\\
&=\frac{1}{2^{n t /2}}\sum_{m,m'}\braket{m|\prod_{\nu=1}^{n} O^{\phantom{\dag}}_\nu O^\dag_\nu|m'} \ket{m}\otimes \ket{m'}^*=\frac{1}{2^{n t /2}}\sum_{m} \ket{m}\otimes \ket{m}^*=\ket{\1}\,.
\end{align}
Here we used \eqref{eq:complexconjugate} and the fact that $O_\nu$ is unitary. Second, we observe that from the definition of $O_\nu$ it directly follows 
\begin{align}
\mathbb P\,\left(\prod_{\nu=1}^{n} O_\nu\otimes O^{*}_\nu\right)\, \mathbb P^{\dag} &= \prod_{\nu=1}^{n} O_\nu\otimes O^{*}_\nu\,. \label{eq:statepropertyi1}
\end{align}
Combining \eqref{eq:statepropertyi1} and \eqref{eq:statepropertyi2} we then have 
\be
\prod_{\nu=1}^{n} O_\nu\otimes O^{*}_\nu \ket{\Psi}= \prod_{\nu=1}^{n} O_\nu\otimes O^{*}_\nu\, \mathbb P\ket{\1}=\mathbb P\, \prod_{\nu=1}^{n} O_\nu\otimes O^{*}_\nu \ket{\1}= \ket{\Psi}\,.
\ee
So we proved \eqref{eq:stateproperty} for any $O_\nu$ acting non trivially, as the unitary operator $O$, only on the $\nu$-th copy of $\mathcal H_t$ in $\mathcal H_{nt}$. The relation \eqref{eq:statepropertyleft} follows immediately by taking the adjoint of
\be
\prod_{\nu=1}^{n} {O^\dag}_\nu\otimes {O^\dag}^{*}_\nu \ket{\Psi}= \ket{\Psi}\,.
\ee

\section{Proof of Property~\eqref{eq:identity}}
\label{app:proofoffund}
In this appendix we prove Eq~\eqref{eq:identity}. 

\begin{proof}
Defining 
\begin{align}
\mathbb J_{\nu,\tau}&= \exp{\left[-i\frac{\pi}{4} \sigma_{\nu,\tau+1}^z\sigma_{\nu,\tau}^z\right]}\otimes\exp{\left[i\frac{\pi}{4} \sigma_{\nu,\tau+1}^z\sigma_{\nu,\tau}^z\right]} & &\tau\in\{1,\dots,t-1\}\,, & &\mathbb J_{\nu,\tau}=\1 & &\tau\leq0,\\
\mathbb Z^h_{\nu,\tau}&= \exp{\left[-i h \sigma_{\nu,\tau}^z\right]}\otimes\exp{\left[i h \sigma_{\nu,\tau}^z\right]} & &\tau\in\{1,\dots,t\}\,, & &\mathbb Z^{h}_{\nu,\tau}=\1 & &\tau\leq0,\\
\mathbb X_{\nu,\tau}&= \exp{\left[i\frac{\pi}{4} \sigma_{\nu,\tau}^x\right]}\otimes\exp{\left[-i\frac{\pi}{4} \sigma_{\nu,\tau}^x\right]} & &\tau\in\{1,\dots,t\}\,, & &\mathbb X_{\nu,\tau}=\1 & &\tau\leq0,
\end{align}
we can rewrite the l.h.s. of \eqref{eq:identity} as follows 
\be
\braket{\Psi|\prod_{j=1}^N \mathbb T_{\tfrac{\pi}{2},\phi_j}[h_j]|\Psi}=\braket{\Psi|\prod_{\nu=1}^n\left[\prod_{j=1}^N\left(\mathbb G^{z}_{\nu, t} \mathbb Z^{\phi_j/2}_{\nu,1}\prod_{\tau=1}^{t-1}\mathbb J_{\nu,\tau} \prod_{\tau=1}^{t}\mathbb Z^{h_j}_{\nu,\tau} \prod_{\tau=1}^{t}\mathbb X_{\nu,\tau}\right)\right]|\Psi}\,.
\ee
To simplify this expression we proceed as follows. First we commute every possible $\mathbb J_{\nu,\tau}, \mathbb Z_{\nu,\tau}, \mathbb X_{\nu,\tau}$ to the left by using the following commutation relations
\begin{align}
\mathbb J^{\phantom{h}}_{\nu,\tau}\mathbb Z^h_{\nu,\tau'}&=\mathbb Z^h_{\nu,\tau'}\mathbb J^{\phantom{h}}_{\nu,\tau} & &\forall \tau,\tau'\,,\\
\mathbb J^{\phantom{h}}_{\nu,\tau}\mathbb X_{\nu,\tau'}&=\mathbb X_{\nu,\tau'}\mathbb J^{\phantom{h}}_{\nu,\tau} & &\tau'\neq\tau,\tau+1\,,  \\
\mathbb X^{\phantom{h}}_{\nu,\tau}\mathbb Z^h_{\nu,\tau'}&=\mathbb Z^h_{\nu,\tau'}\mathbb X^{\phantom{h}}_{\nu,\tau} & &\tau'\neq\tau\,,\\
\mathbb J^{\phantom{h}}_{\nu,\tau}\mathbb G^{z}_{\nu, \tau'}&=\mathbb G^{z}_{\nu, \tau'}\mathbb J^{\phantom{h}}_{\nu,\tau} & &\forall \tau',\tau\,,\\
\mathbb Z^{h}_{\nu,\tau}\mathbb G^{z}_{\nu, \tau'}&=\mathbb G^{z}_{\nu, \tau'}\mathbb Z^{h}_{\nu,\tau} & &\forall \tau',\tau\,,\\
\mathbb X^{\phantom{h}}_{\nu,\tau}\mathbb G^{z}_{\nu, \tau'}&=\mathbb G^{z}_{\nu, \tau'}\mathbb X^{\phantom{h}}_{\nu,\tau} & &\tau'\neq\tau\,.
\end{align}
Then we use  
\begin{align}
\bra{\Psi}\prod_{\nu=1}^n\mathbb J^{\phantom{h}}_{\nu,\tau}&=\bra{\Psi}\,,\label{eq:Mstateleft}\\
\bra{\Psi}\prod_{\nu=1}^n\mathbb X_{\nu,\tau'}&=\bra{\Psi}\,,  \\
\bra{\Psi}\prod_{\nu=1}^n\mathbb Z^h_{\nu,\tau'}&=\bra{\Psi}\,,
\end{align}
which follow from \eqref{eq:stateproperty}. Finally, using also 
\begin{align}
\prod_{\nu=1}^n\mathbb J^{\phantom{h}}_{\nu,\tau}\ket{\Psi}&=\ket{\Psi}\,,\label{eq:Jstateright}\\
\prod_{\nu=1}^n\mathbb X_{\nu,\tau'}\ket{\Psi}&=\ket{\Psi}\,, \label{eq:Xstateright} \\
\prod_{\nu=1}^n \mathbb Z^h_{\nu,\tau'}\ket{\Psi}&=\ket{\Psi}\,,\label{eq:Zstateright}
\end{align}
following from \eqref{eq:statepropertyleft}, on the rightmost term in the product over $j$ we find   
\be
\braket{\Psi|\prod_{j=1}^N \mathbb T_{\tfrac{\pi}{4},\phi_j}[h_j]|\Psi}=\braket{\Psi|\prod_{\nu=1}^n\left[\prod_{j=1}^{N-1} \mathbb A_{\nu,j}\right] |\Psi}\,,
\ee
where we defined 
\begin{align}
\mathbb A_{\nu,j}&=\mathbb G^{z}_{\nu, t} \!\!\!\!\prod_{\tau=t-j+1}^{t-1}\!\!\!\!\mathbb J_{\nu,\tau}\!\!\!\! \prod_{\tau=t-j+2}^{t}\!\!\!\!\tilde{\mathbb Z}^{h_j,\phi_j}_{\nu,\tau}\!\!\!\! \prod_{\tau=t-j+1}^{t}\!\!\!\! \mathbb X_{\nu,\tau}\,\, \mathbb G^{z}_{\nu, t}\,,
\label{eq:Adef}\\
\tilde{\mathbb{Z}}^{h,\phi}_{\nu,\tau} &=\mathbb Z^{h+(\phi/2) \delta_{\tau,1}}_{\nu,\tau}\,.
\end{align}
Using 
\begin{align}
\mathbb G^{z}_{\nu, \tau} \mathbb X^{\phantom{z}}_{\nu, \tau} \mathbb G^{z}_{\nu, \tau}&=\mathbb G^{z}_{\nu, \tau} \left[\mathbb G^{x}_{\nu, \tau}+\frac{i}{2} \left[\sigma_{\nu,\tau}^x\otimes\1-\1\otimes\sigma_{\nu,\tau}^x\right] \right]\mathbb G^{z}_{\nu, \tau}\notag\\
&=\mathbb G^{z}_{\nu, \tau} \left[\mathbb G^{z}_{\nu, \tau} \mathbb G^{x}_{\nu, \tau}+\frac{i}{4} \left[\1 - \sigma^z_{\nu,\tau} \otimes \sigma^z_{\nu,\tau}\right] \left[\sigma_{\nu,\tau}^x\otimes\1-\1\otimes\sigma_{\nu,\tau}^x\right] \right]\notag\\
&=\mathbb G^{z}_{\nu, \tau}\mathbb G^{x}_{\nu, \tau}\mathbb G^{z}_{\nu, \tau}=\mathbb G^{z}_{\nu, \tau}\mathbb G^{x}_{\nu, \tau}\,,
\label{eq:usefulrel1}
\end{align}
we can rewrite \eqref{eq:Adef} as follows 
\be
\mathbb A_{\nu,j}= \mathbb G^{z}_{\nu, t}\mathbb J_{\nu,t-1}\mathbb G^{x}_{\nu, t} \!\!\!\!\prod_{\tau=t-j+1}^{t-2}\!\!\!\!\mathbb J_{\nu,\tau}\!\!\!\! \prod_{\tau=t-j+2}^{t-1}\!\!\!\!\tilde{\mathbb Z}^{h_j,\phi_j}_{\nu,\tau}\!\!\!\! \prod_{\tau=t-j+1}^{t-1}\!\!\!\! \mathbb X_{\nu,\tau}. 
\label{eq:Adefsimp}
\ee
We now make use the following Lemma, proven in Appendix~\ref{app:prooflemma}, to simplify the products of $\mathbb A_{\nu,j}$s 
\begin{lemma} 
\label{lemmaA}
\begin{align}
\mathbb A_{\nu,1}\cdots \mathbb A_{\nu,2n} &=\mathbb J_{\nu,t-1}\prod_{j=0}^{n-1}\left[\mathbb G^{z}_{\nu, t-j}\mathbb G^{x}_{\nu, t-j}\right]\mathbb G^{z}_{\nu, t-n} \mathbb X^{\phantom{z}}_{\nu,t-n} \prod_{j=n}^{2n-2}\left[\prod_{\tau=t-1-j}^{t-2n+j}\!\!\!\!\!\!  \mathbb J_{\nu,\tau} \!\! \prod_{\tau=t-j}^{t-2n+j}\!\!\!\tilde{\mathbb Z}^{h_{j+2},\phi_{j+2}}_{\nu,\tau} \!\!\prod_{\tau=t-1-j}^{t-2n+j+1}\!\!\!\!\!\!\mathbb X_{\nu,\tau}\right]\!\!,\label{eq:indreleven}\\
\mathbb A_{\nu,1}\cdots \mathbb A_{\nu,2n+1} &=\mathbb J_{\nu,t-1}\prod_{j=0}^{n}\left[\mathbb G^{z}_{\nu, t-j}\mathbb G^{x}_{\nu, t-j}\right] \prod_{j=n}^{2n-1}\left[\prod_{\tau=t-1-j}^{t-2n-1+j}\!\!\!\!\!\!  \mathbb J_{\nu,\tau} \!\! \prod_{\tau=t-j}^{t-2n+j-1}\!\!\!\!\tilde{\mathbb Z}^{h_{j+2},\phi_{j+2}}_{\nu,\tau}\!\!\prod_{\tau=t-1-j}^{t-2n+j}\!\!\!\!\!\!\mathbb X_{\nu,\tau}\right]\!\!,\qquad\qquad n\geq1\,. \label{eq:indrelodd}
\end{align}
\end{lemma}
Using now \eqref{eq:Mstateleft}, \eqref{eq:Jstateright}, \eqref{eq:Xstateright}, and \eqref{eq:Zstateright} we have 
\be
\braket{\Psi|\prod_{j=1}^N \mathbb T_{\tfrac{\pi}{4},\phi_j}[h_j] |\Psi}=\braket{\Psi|\prod_{\nu=1}^{n}\left[\prod_{j=0}^{\lfloor \frac{N}{2}\rfloor-1}\!\!\!\!\!\left[\mathbb G^{z}_{\nu, t-j}\mathbb G^{x}_{\nu, t-j}\right][\mathbb G^{z}_{\nu, t-\lfloor N/2\rfloor}]^{{\rm mod}(N,2)}\right]|\Psi}\,,
\ee
which concludes the proof. 
\end{proof}

\subsection{Proof of Lemma~\ref{lemmaA}.}
\label{app:prooflemma}
 
Here we prove Lemma~\ref{lemmaA}. 
 
\begin{proof}
We proceed by induction in the number of terms in the products of $\mathbb A_{\nu,j}$s. First we establish the basis. We begin by computing 
\begin{align}
\mathbb A_{\nu,1} \mathbb A_{\nu,2}&= \mathbb G^{z}_{\nu, t}\mathbb J_{\nu,t-1}\mathbb G^{x}_{\nu, t}\mathbb G^{z}_{\nu, t}\mathbb J_{\nu,t-1}\mathbb G^{x}_{\nu, t} \mathbb X_{\nu,t-1} \notag\\
&=\mathbb J_{\nu,t-1} \mathbb G^{z}_{\nu, t}\mathbb G^{x}_{\nu, t}\mathbb G^{z}_{\nu, t-1} \mathbb X^{\phantom{z}}_{\nu,t-1}\,.
\label{eq:basiseven}
\end{align}
where we used 
\begin{align}
\mathbb G^{z}_{\nu, \tau}  \mathbb G^{x}_{\nu, \tau} \mathbb J^{\phantom{z}}_{\nu,\tau-1} \mathbb G^{x}_{\nu, \tau}&=\mathbb G^{z}_{\nu, \tau} \mathbb G^{x}_{\nu, \tau} \left[\mathbb G^{z}_{\nu, \tau-1}+\frac{i}{2} \left[\sigma_{\nu,\tau-1}^z\otimes \sigma_{\nu,\tau}^z - \sigma_{\nu,\tau}^z \otimes\sigma_{\nu,\tau-1}^z\right] \right]\mathbb G^{x}_{\nu, t}\notag\\
&=\mathbb G^{z}_{\nu, \tau} \mathbb G^{x}_{\nu, \tau} \left[\mathbb G^{x}_{\nu, \tau} \mathbb G^{z}_{\nu, \tau-1}+\frac{i}{4} \left[\1 - \sigma^x_{\nu,\tau} \otimes \sigma^x_{\nu,\tau}\right] \left[\sigma_{\nu,\tau-1}^z\otimes \sigma_{\nu,\tau}^z - \sigma_{\nu,\tau}^z \otimes\sigma_{\nu,\tau-1}^z\right] \right]\notag\\
&=\mathbb G^{z}_{\nu, \tau}\mathbb G^{x}_{\nu, \tau}\mathbb G^{z}_{\nu, \tau-1}\,.
\label{eq:usefulrel2}
\end{align}
We see that \eqref{eq:basiseven} agrees with \eqref{eq:indreleven} for $n=1$. We then compute  
\begin{align}
\mathbb A_{\nu,1} \mathbb A_{\nu,2} \mathbb A_{\nu,3} &=\mathbb G^{z}_{\nu, t}\mathbb J_{\nu,t-1}\mathbb G^{x}_{\nu, t}\mathbb G^{z}_{\nu, t}\mathbb J_{\nu,t-1}\mathbb G^{x}_{\nu, t} \mathbb X_{\nu,t-1}\mathbb G^{z}_{\nu, t}\mathbb J_{\nu,t-1}\mathbb G^{x}_{\nu, t} \mathbb J_{\nu,t-2} \tilde {\mathbb Z}^{h_{3},\phi_{3}}_{\nu,t-1}\mathbb X_{\nu,t-2}\mathbb X_{\nu,t-1}\notag\\
&=\mathbb J_{\nu,t-1} \mathbb G^{z}_{\nu, t}\mathbb G^{x}_{\nu, t}\mathbb G^{z}_{\nu, t-1} \mathbb X^{\phantom{z}}_{\nu,t-1}\mathbb G^{z}_{\nu, t} \mathbb J_{\nu,t-1}\mathbb G^{x}_{\nu, t} \mathbb J_{\nu,t-2} \tilde{\mathbb Z}^{h_{3},\phi_{3}}_{\nu,t-1}\mathbb X_{\nu,t-2}\mathbb X_{\nu,t-1}\notag\\
&=\mathbb J_{\nu,t-1} \mathbb G^{z}_{\nu, t}\mathbb G^{x}_{\nu, t}\mathbb G^{z}_{\nu, t-1} \mathbb X^{\phantom{z}}_{\nu,t-1}\mathbb G^{z}_{\nu, t-1} \mathbb J_{\nu,t-2} \tilde{\mathbb Z}^{h_{3},\phi_{3}}_{\nu,t-1}\mathbb X_{\nu,t-2}\mathbb X_{\nu,t-1}\notag\\
&=\mathbb J_{\nu,t-1} \mathbb G^{z}_{\nu, t}\mathbb G^{x}_{\nu, t}\mathbb G^{z}_{\nu, t-1} \mathbb G^{x}_{\nu, t-1} \mathbb J_{\nu,t-2} \tilde{\mathbb Z}^{h_{3},\phi_{3}}_{\nu,t-1} \mathbb X_{\nu,t-2}\mathbb X_{\nu,t-1}\notag\\
&=\mathbb J_{\nu,t-1} \mathbb G^{z}_{\nu, t}\mathbb G^{x}_{\nu, t}\mathbb G^{z}_{\nu, t-1} \mathbb G^{x}_{\nu, t-1} \mathbb J_{\nu,t-2} \mathbb X_{\nu,t-2}\mathbb X_{\nu,t-1}\,.
\label{eq:basisodd}
\end{align}
In the last step we used 
\be
\mathbb G^{z}_{\nu, \tau} \tilde{\mathbb Z}^{h,\phi}_{\nu,\tau}=\mathbb G^{z}_{\nu, \tau}\,.
\ee
Since \eqref{eq:basisodd} agrees with \eqref{eq:indrelodd} for $n=1$ we successfully established the basis for the inductive procedure. To conclude we need to prove that
\begin{itemize}
\item[$(i)$] if \eqref{eq:indreleven} holds for $n$ then \eqref{eq:indrelodd} holds for $n$
\item[$(ii)$] if \eqref{eq:indrelodd} holds for $n$ then \eqref{eq:indreleven} holds for $n+1$.   
\end{itemize}
Let us prove $(i)$. Assuming \eqref{eq:indreleven} we have 
\begin{align}
\mathbb A_{\nu,1}\cdots \mathbb A_{\nu,2n}\mathbb A_{\nu,2n+1} =& \mathbb J_{\nu,t-1}\prod_{j=0}^{n-1}\left[\mathbb G^{z}_{\nu, t-j}\mathbb G^{x}_{\nu, t-j}\right]\mathbb G^{z}_{\nu, t-n} \mathbb X^{\phantom{z}}_{\nu,t-n} \prod_{j=n}^{2n-2}\left[\prod_{\tau=t-1-j}^{t-2n+j}\!\!\!\!\!\!  \mathbb J_{\nu,\tau} \!\! \prod_{\tau=t-j}^{t-2n+j}\!\!\!\tilde{\mathbb Z}^{h_{j+2},\phi_{j+2}}_{\nu,\tau} \!\!\prod_{\tau=t-1-j}^{t-2n+j+1}\!\!\!\!\!\!\mathbb X_{\nu,\tau}\right]\!\!,\notag\\
&\times \mathbb G^{z}_{\nu, t}\mathbb J_{\nu,t-1}\mathbb G^{x}_{\nu, t} \!\!\!\!\prod_{\tau=t-2n}^{t-2}\!\!\!\!\mathbb J_{\nu,\tau}\!\!\!\! \prod_{\tau=t-2n+1}^{t-1}\!\!\!\!\tilde{\mathbb Z}^{h_{2n+1},\phi_{2n+1}}_{\nu,\tau}\!\!\!\! \prod_{\tau=t-2n}^{t-1}\!\!\!\! \mathbb X_{\nu,\tau}\notag\\
=& \mathbb J_{\nu,t-1}\prod_{j=0}^{n-1}\left[\mathbb G^{z}_{\nu, t-j}\mathbb G^{x}_{\nu, t-j}\right]\mathbb G^{z}_{\nu, t-n} \mathbb X^{\phantom{z}}_{\nu,t-n} \prod_{j=n}^{2n-2}\left[\prod_{\tau=t-1-j}^{t-2n+j}\!\!\!\!\!\!  \mathbb J_{\nu,\tau} \!\! \prod_{\tau=t-j}^{t-2n+j}\!\!\!\tilde {\mathbb Z}^{h_{j+2},\phi_{j+2}}_{\nu,\tau} \!\!\prod_{\tau=t-1-j}^{t-2n+j+1}\!\!\!\!\!\!\mathbb X_{\nu,\tau}\right]\!\!,\notag\\
&\times \mathbb G^{z}_{\nu, t-1} \!\!\!\!\prod_{\tau=t-2n}^{t-2}\!\!\!\!\mathbb J_{\nu,\tau}\!\!\!\! \prod_{\tau=t-2n+1}^{t-1}\!\!\!\!\tilde{\mathbb Z}^{h_{2n+1},\phi_{2n+1}}_{\nu,\tau}\!\!\!\! \prod_{\tau=t-2n}^{t-1}\!\!\!\! \mathbb X_{\nu,\tau}\notag\\
\end{align}
where we used that $\mathbb G^{x}_{\nu, t}$ commutes with all the terms on the first line to bring it close to $\mathbb J_{\nu,t-1}$. Then we employed \eqref{eq:usefulrel2}. Moving the projector $\mathbb G^{z}_{\nu, t-1}$ on the second line to the left and using multiple times \eqref{eq:usefulrel1} and \eqref{eq:usefulrel2} we have 
\begin{align}
\mathbb A_{\nu,1}\cdots \mathbb A_{\nu,2n}\mathbb A_{\nu,2n+1} =& \mathbb J_{\nu,t-1}\prod_{j=0}^{n-1}\left[\mathbb G^{z}_{\nu, t-j}\mathbb G^{x}_{\nu, t-j}\right]\mathbb G^{z}_{\nu, t-n} \mathbb X^{\phantom{z}}_{\nu,t-n} \mathbb G^{z}_{\nu, t-n}  \prod_{j=n}^{2n-2}\left[\prod_{\tau=t-1-j}^{t-2n+j-1}\!\!\!\!\!\!  \mathbb J_{\nu,\tau} \!\! \prod_{\tau=t-j}^{t-2n+j-1}\!\!\!\tilde{\mathbb Z}^{h_{j+2},\phi_{j+2}}_{\nu,\tau} \!\!\prod_{\tau=t-1-j}^{t-2n+j}\!\!\!\!\!\!\mathbb X_{\nu,\tau}\right]\!\!,\notag\\
&\times \!\!\!\!\prod_{\tau=t-2n}^{t-2}\!\!\!\!\mathbb J_{\nu,\tau}\!\!\!\! \prod_{\tau=t-2n+1}^{t-1}\!\!\!\!\tilde{\mathbb Z}^{h_{2n+1},\phi_{2n+1}}_{\nu,\tau}\!\!\!\! \prod_{\tau=t-2n}^{t-1}\!\!\!\! \mathbb X_{\nu,\tau}\notag\\
=& \mathbb J_{\nu,t-1}\prod_{j=0}^{n-1}\left[\mathbb G^{z}_{\nu, t-j}\mathbb G^{x}_{\nu, t-j}\right]\mathbb G^{z}_{\nu, t-n}  \mathbb G^{x}_{\nu, t-n}  \prod_{j=n}^{2n-2}\left[\prod_{\tau=t-1-j}^{t-2n+j-1}\!\!\!\!\!\!  \mathbb J_{\nu,\tau} \!\! \prod_{\tau=t-j}^{t-2n+j-1}\!\!\!\tilde{\mathbb Z}^{h_{j+2},\phi_{j+2}}_{\nu,\tau} \!\!\prod_{\tau=t-1-j}^{t-2n+j}\!\!\!\!\!\!\mathbb X_{\nu,\tau}\right]\!\!,\notag\\
&\times \!\!\!\!\prod_{\tau=t-2n}^{t-2}\!\!\!\!\mathbb J_{\nu,\tau}\!\!\!\! \prod_{\tau=t-2n+1}^{t-1}\!\!\!\!\tilde{\mathbb Z}^{h_{2n+1},\phi_{2n+1}}_{\nu,\tau}\!\!\!\! \prod_{\tau=t-2n}^{t-1}\!\!\!\! \mathbb X_{\nu,\tau}\notag\\
=& \mathbb J_{\nu,t-1}\prod_{j=0}^{n}\left[\mathbb G^{z}_{\nu, t-j}\mathbb G^{x}_{\nu, t-j}\right] \prod_{j=n}^{2n-1}\left[\prod_{\tau=t-1-j}^{t-2n+j-1}\!\!\!\!\!\!  \mathbb J_{\nu,\tau} \!\! \prod_{\tau=t-j}^{t-2n+j-1}\!\!\!\tilde{\mathbb Z}^{h_{j+2},\phi_{j+2}}_{\nu,\tau} \!\!\prod_{\tau=t-1-j}^{t-2n+j}\!\!\!\!\!\!\mathbb X_{\nu,\tau}\right]\!\!,
\end{align}
which is exactly \eqref{eq:indrelodd}. Let us now prove $(ii)$. Assuming \eqref{eq:indrelodd} we have 
\begin{align}
\mathbb A_{\nu,1}\cdots \mathbb A_{\nu,2n+1}\mathbb A_{\nu,2n+2} =& \mathbb J_{\nu,t-1}\prod_{j=0}^{n}\left[\mathbb G^{z}_{\nu, t-j}\mathbb G^{x}_{\nu, t-j}\right] \prod_{j=n}^{2n-1}\left[\prod_{\tau=t-1-j}^{t-2n-1+j}\!\!\!\!\!\!  \mathbb J_{\nu,\tau} \!\! \prod_{\tau=t-j}^{t-2n+j-1}\!\!\!\!\tilde{\mathbb Z}^{h_{j+2},\phi_{j+2}}_{\nu,\tau}\!\!\prod_{\tau=t-1-j}^{t-2n+j}\!\!\!\!\!\!\mathbb X_{\nu,\tau}\right]\notag\\
&\times  \mathbb G^{z}_{\nu, t}\mathbb J_{\nu,t-1}\mathbb G^{x}_{\nu, t} \!\!\!\!\prod_{\tau=t-2n-1}^{t-2}\!\!\!\!\mathbb J_{\nu,\tau}\!\!\!\! \prod_{\tau=t-2n}^{t-1}\!\!\!\!\tilde{\mathbb Z}^{h_{2n+2},\phi_{2n+2}}_{\nu,\tau}\!\!\!\! \prod_{\tau=t-2n-1}^{t-1}\!\!\!\! \mathbb X_{\nu,\tau}\notag\\
=& \mathbb J_{\nu,t-1}\prod_{j=0}^{n}\left[\mathbb G^{z}_{\nu, t-j}\mathbb G^{x}_{\nu, t-j}\right] \prod_{j=n}^{2n-1}\left[\prod_{\tau=t-1-j}^{t-2n-1+j}\!\!\!\!\!\!  \mathbb J_{\nu,\tau} \!\! \prod_{\tau=t-j}^{t-2n+j-1}\!\!\!\!\tilde{\mathbb Z}^{h_{j+2},\phi_{j+2}}_{\nu,\tau}\!\!\prod_{\tau=t-1-j}^{t-2n+j}\!\!\!\!\!\!\mathbb X_{\nu,\tau}\right]\notag\\
&\times  \mathbb G^{z}_{\nu, t-1} \!\!\!\!\prod_{\tau=t-2n-1}^{t-2}\!\!\!\!\mathbb J_{\nu,\tau}\!\!\!\! \prod_{\tau=t-2n}^{t-1}\!\!\!\!\tilde{\mathbb Z}^{h_{2n+2},\phi_{2n+2}}_{\nu,\tau}\!\!\!\! \prod_{\tau=t-2n-1}^{t-1}\!\!\!\! \mathbb X_{\nu,\tau}
\end{align}
where we again used that $\mathbb G^{x}_{\nu, t}$ commutes with all the terms on the first line to bring it close to $\mathbb J_{\nu,t-1}$ and employed \eqref{eq:usefulrel2}. Moving now the projector $\mathbb G^{z}_{\nu, t-1}$ on the second line to the left and using \eqref{eq:usefulrel1} and \eqref{eq:usefulrel2} we have
\begin{align}
\mathbb A_{\nu,1}\cdots \mathbb A_{\nu,2n+2} =& \mathbb J_{\nu,t-1}\prod_{j=0}^{n}\left[\mathbb G^{z}_{\nu, t-j}\mathbb G^{x}_{\nu, t-j}\right]  \mathbb G^{z}_{\nu, t-n-1}\mathbb X^{\phantom{z}}_{\nu,t-n-1}  \prod_{j=n}^{2n-1}\left[\prod_{\tau=t-1-j}^{t-2n-2+j}\!\!\!\!\!\!  \mathbb J_{\nu,\tau} \!\! \prod_{\tau=t-j}^{t-2n+j-2}\!\!\!\!\tilde{\mathbb Z}^{h_{j+2},\phi_{j+2}}_{\nu,\tau}\!\!\!\!\!\prod_{\tau=t-1-j}^{t-2n-1+j}\!\!\!\!\!\!\mathbb X_{\nu,\tau}\right]\notag\\
&\times \!\!\!\!\prod_{\tau=t-2n-1}^{t-2}\!\!\!\!\mathbb J_{\nu,\tau}\!\!\!\! \prod_{\tau=t-2n}^{t-1}\!\!\!\!\tilde{\mathbb Z}^{h_{2n+2},\phi_{2n+2}}_{\nu,\tau}\!\!\!\! \prod_{\tau=t-2n-1}^{t-1}\!\!\!\! \mathbb X_{\nu,\tau}\notag\\
=& \mathbb J_{\nu,t-1}\prod_{j=0}^{n}\left[\mathbb G^{z}_{\nu, t-j}\mathbb G^{x}_{\nu, t-j}\right]  \mathbb G^{z}_{\nu, t-n-1}\mathbb X^{\phantom{z}}_{\nu,t-n-1}  \prod_{j=n}^{2n}\left[\prod_{\tau=t-1-j}^{t-2n-2+j}\!\!\!\!\!\!  \mathbb J_{\nu,\tau} \!\! \prod_{\tau=t-j}^{t-2n+j-2}\!\!\!\!\tilde{\mathbb Z}^{h_{j+2},\phi_{j+2}}_{\nu,\tau}\!\!\!\!\!\prod_{\tau=t-1-j}^{t-2n-1+j}\!\!\!\!\!\!\mathbb X_{\nu,\tau}\right]\!\!,
\end{align} 
which is exactly \eqref{eq:indreleven} for $n+1$. This concludes the proof. 

\end{proof}

\end{widetext}


\begin{thebibliography}{99}
\bibitem{note2}
Note that the notions ``quantum chaotic" and ``ergodic" are equivalent for a quantum many-body system and are used interchangeably in the paper.
\bibitem{note3}
Note that for $\theta_j\in\{0,\pi\}$ the states \eqref{eq:state0} are independent of $\boldsymbol \phi$. This means that we do not need to specify the value of $\boldsymbol \phi$ in \eqref{eq:classesL}.
\bibitem{notegaussian}
Note that the density of free (Bogoliubov) fermions is independent of $\phi$. The $\phi$-dependence of $S^{(n)}_{\theta,\phi}$  (see the inset of Fig.~\ref{fig:plotIntegrable}) then implies that the latter is not entirely fixed by the fermionic density. This is expected for non-Gaussian states.
\bibitem{notepowermethod}
Note that $m=J_{\rm max}-1$, where $J_{\rm max}$ is the size of the largest Jordan block associated to the eigenvalue 0. For separating states we have ${m=2t}$. Our numerical analysis shows that this remains true for all states \eqref{eq:state0}.
\end{thebibliography}
\end{document}